\documentclass[apj,numbers,numberedappendix]{emulateapj}
\usepackage{graphicx,  bm, color, amsmath, amssymb, xfrac}

\usepackage[usenames,dvipsnames]{xcolor}
\def\corr{}

\usepackage{hyperref}
\hypersetup{
    colorlinks = true,
    citecolor = {MidnightBlue},
    linkcolor = {BrickRed},
    urlcolor = {BrickRed}
}

\def\aH{\mathcal{H}}
\def\be{\begin{equation}}
\def\ee{\end{equation}}
\def\bea{\begin{eqnarray}}
\def\eea{\end{eqnarray}}
\newcommand{\Hu}{{\cal H}}
\newcommand{\p}{^{\prime}}
\newcommand{\pp}{^{\prime\prime}}
\newcommand{\mt}{{\tilde\mu}}

\newcommand{\cR}{{\cal R}}

\newcommand{\coR}{{\cal R_{\mathrm{co}}}}
\newcommand{\A}{{\frac{9}{2}\frac{\Hu^2}{k^2}\Omega_{\rm M}}}

\begin{document}

\title{Observational signatures of modified gravity on ultra-large scales}

\author{Tessa Baker}
\email{tessa.baker@physics.ox.ac.uk}
\affil{Dept. of Astrophysics, University of Oxford, Denys Wilkinson Building, Keble Road, Oxford, OX1 3RH, UK}
\author{Philip Bull}
\email{p.j.bull@astro.uio.no}
\affil{Institute of Theoretical Astrophysics, University of Oslo, P.O. Box 1029 Blindern,
       N-0315 Oslo, Norway}

\begin{abstract}
Extremely large surveys with future experiments like Euclid and the SKA will soon allow us to access perturbation modes close to the Hubble scale, with wavenumbers $k \sim \aH$. If a modified gravity theory is responsible for cosmic acceleration, the Hubble scale is a natural regime for deviations from General Relativity (GR) to become manifest. The majority of studies to date have concentrated on the consequences of alternative gravity theories for the subhorizon, quasi-static regime, however. {\corr In this paper} we investigate how modifications to the gravitational field equations affect perturbations around the Hubble scale. {\corr We choose functional forms to represent the generic scale-dependent behaviour of gravity theories that modify GR at long wavelengths, and study the resulting deviations of ultra large-scale relativistic observables from their GR behaviour. We find that these are small unless modifications to the field equations are drastic.} The angular dependence and redshift evolution of the deviations is highly parameterisation- and survey-dependent, however, and so they are possibly a rich source of modified gravity phenomenology if they can be measured.
\end{abstract}


\section{Introduction}
\label{sec:intro}

The past decade of work on extensions of General Relativity (GR) has primarily been motivated by the hope of explaining cosmic acceleration without a fine-tuned cosmological constant, $\Lambda$. The inferred energy scale of $\Lambda$ is approximately the Hubble scale today, $k \sim H_0 \approx 2 \times 10^{-4}$ Mpc$^{-1}$; so, for theories hoping to dispose of $\Lambda$, it seems natural that the cosmological horizon should act as a threshold at which corrections to GR become relevant. Studying how modified gravity (MG) theories affect Hubble-scale perturbation modes could therefore offer interesting insights into the dark energy and cosmological constant problems.

Long wavelength (infrared; IR) modifications of gravity have mostly been explored in the context of braneworld models \citep{DGO2000, 2004LRR.....7....7M}, non-local theories {\corr \citep{Maggiore2014,Dirian2014, Conroy2015}} and massive gravity/bigravity \citep{deRham:2010kj, 2011PhRvD..84l4046D, Hassan:2011hr}. In some theories the features causing the long-wavelength modifications also result in ultraviolet (UV) phenomenology, such as the modification of forces at a source-dependent distance $r_*$ \citep{Babichev2013}, extra graviton polarisations \citep{Dvali2004}, and the existence of a UV cut-off to the theory. In this paper we will focus solely on the IR aspects of modifications to gravity; any associated UV phenomenology is unlikely to affect the observables we consider here.

In contrast to the demanding simulation requirements and astrophysical uncertainties that plague non-linear wave modes $k\gtrsim 0.1$ Mpc$^{-1}$, ultra-large scales are a relatively `clean' and safely linear regime in which to test gravity. Perturbation modes with wavelengths approaching the Hubble scale are mostly beyond the reach of current large-scale structure experiments however \citep[although see, e.g. ][]{2013MNRAS.435.1857L, 2014MNRAS.438.1724H}, and so the bulk of work on constraining alternative gravity theories has focussed on significantly subhorizon scales, often using quasi-static approximations to simplify calculations \citep{Amendola:2012hg,Silvestri2013,2015arXiv150306831S}. The observational situation is soon set to change though, as forthcoming surveys of large-scale structure -- notably wide-angle galaxy and 21cm intensity mapping surveys \citep{2009arXiv0912.0201L, 2011arXiv1106.1706S, 2013LRR....16....6A, Bull:2014vw, 2015arXiv150103825J} -- will make it possible to probe ultra-large scales across gigantic survey volumes at $z \lesssim 5$ \citep{2012PhRvD..86f3514Y, Camera:2013kpa, Camera2015, 2015arXiv150506179R, Alonso2015}.

The purpose of this article is to understand how Hubble-scale gravitational corrections affect various observables on ultra-large scales. The space of MG theories is complex, and there is no consensus on a favoured model. {\corr Hence, as far as possible, we will draw only upon generic physical features (Lorentz invariance, mass scales, etc.) instead of adopting a particular theory. From \S\ref{subsec:example} onwards we choose some specific ansatzes in order to obtain example numerical results; our calculations up to this point are general, however.}

\begin{figure}[t]
\vspace{-0.5em}\hspace{-1.1em}\includegraphics[width=1.09\columnwidth]{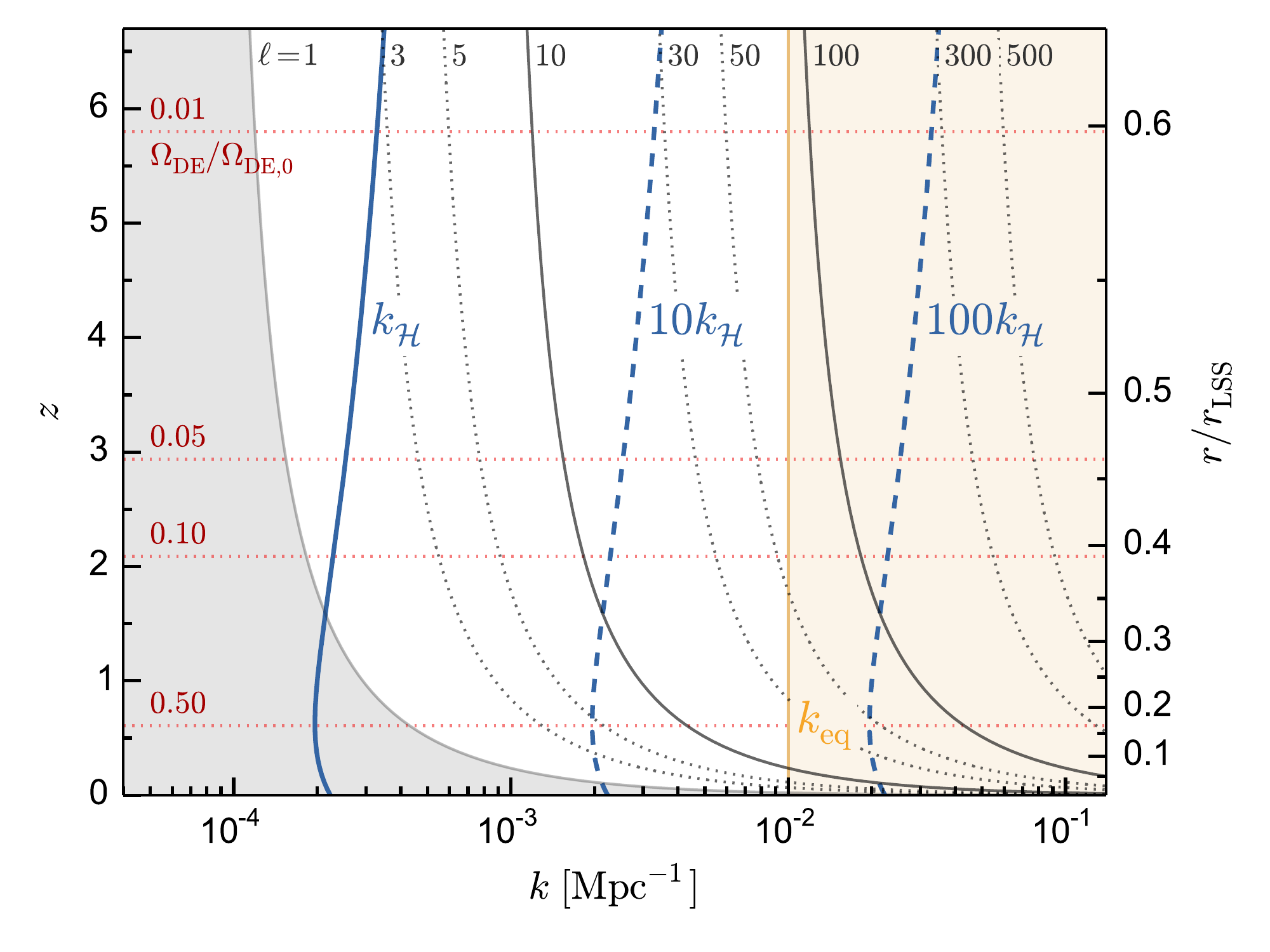}\vspace{-1em}
\caption{Comoving Hubble scale, $k_\aH = \aH$, as a function of redshift and comoving distance to the last scattering surface (blue solid/dashed lines). Also shown are the (Limber-approximated) comoving scales probed by a given multipole, $\ell \simeq k r$ (black solid/dotted lines), the matter-radiation equality scale, $k_{\rm eq}$ (yellow line), and the normalised fractional energy density in dark energy, $\Omega_{\rm DE}(z)/\Omega_{\rm DE, 0}$ (horizontal red lines). \citep[c.f.][]{2014arXiv1409.3831Z}}
\label{fig:horizon}\vspace{-0.4em}
\end{figure}

Horizon-scale modes require careful handling: calculations performed in different gauges will no longer agree in general\footnote{{\corr Unless the quantity being calculated is a true observable.}}, and relativistic effects become important. The relativistic corrections impacting horizon-scale large-scale structure observables in GR were worked out in \cite{2009PhRvD..80h3514Y, 2010PhRvD..82h3508Y, 2011PhRvD..84f3505B, 2011PhRvD..84d3516C, 2012PhRvD..86f3514Y, Jeong:2011as}; in \S\ref{sec:CLASS} we will consider how they are affected by modifications to the gravitational field equations. Detection of the relativistic effects will be challenging \citep{2012PhRvD..86f3514Y, 2013JCAP...02..044M, Alonso2015}, as their impact is limited to low multipoles of the angular power spectrum of the galaxy distribution (see Fig.~\ref{fig:horizon}). A number of other physical and systematic effects can also contaminate the signal \citep{2012PhRvD..85d1301B, 2013PASP..125..705P, 2013MNRAS.435.1857L, Alonso2015}. Nevertheless, it is important to take the relativistic corrections into account when studying the Hubble scale.

This paper is structured as follows. After motivating our effective description of the gravitational field equations (\S\ref{sec:params}), we first build intuition about how the dark matter density perturbation is altered by horizon-scale deviations from GR (\S\ref{sec:set-up}). We then apply these ideas to a large-scale structure observable, the angular power spectrum of source number counts (\S\ref{sec:CLASS}), adopting a formulation that can describe both number counts and brightness temperature fluctuations simultaneously. We discuss our results and their observational feasibility in \S\ref{sec:conclusions}.

\section{Parameterisation} 
\label{sec:params}

Observations to date are consistent with a constant dark energy equation of state of $w=-1$ \citep{Planck:2015tt}, leaving the evolution of perturbations as the most likely arena for seeing deviations from $\Lambda$CDM on cosmological scales. As such, we will assume a cosmological expansion history that is indistinguishable from $\Lambda$CDM in what follows. This restriction of a fixed background can easily be relaxed at the expense of introducing more parameters into our framework.

The ultra-large scale modes that we are interested in are very well-described by linear perturbation theory. Studies focussing on the quasi-static regime have shown that gravity theories endowed with a single scalar degree of freedom are encapsulated by the following effective description on small (but still linear) scales (see \S\ref{subsec:defs} for our conventions regarding the metric potentials):\footnote{Eq.~(\ref{Poisson_gi}) is not strictly the Poisson equation, as it has the timelike gravitational potential on the LHS (i.e.~$\Psi$ instead of $\Phi$).}
\bea
-2k^2\hat\Psi&=&8\pi G_{\!N} a^2 \rho D \mt\left(a, k\right) \label{Poisson_gi}\\
\hat\Phi&=&\hat\Psi \, \gamma\left(a, k\right), \label{slip_gi}
\eea
where $D$ is a gauge-invariant fractional density perturbation, and a sum over fluid species is implied on the RHS of Eq.~(\ref{Poisson_gi}). (Note that $D$ here is often denoted by $\Delta$ in other works; we follow the conventions of \cite{Durrer2005} used by the CLASS code in \S\ref{sec:CLASS}). 
The hats on the metric potentials indicate fully gauge-invariant quantities (the Bardeen potentials), which we introduce for later convenience. $\left\{\mt,\gamma\right\}$ are two functions of conformal time and scale. For theories with second-order equations of motion (e.o.m.s), their $k$-dependence is known; it has been derived for particular models in \citet{deFelice,Gleyzes2013,Bloomfield2013}, but is in fact a straightforward consequence of a few general principles: gauge invariance, Lorentz invariance, and second order e.o.m.s. \citep{Baker:2014tc}.

With a little care, the same arguments can be used to justify the use of Eqs.~(\ref{Poisson_gi}) and (\ref{slip_gi}) on scales beyond the quasi-static regime too. The scale-dependence of $\mt$ and $\gamma$ is a little more complex here, as terms that go like $\Hu/k$ cannot be neglected. Similarly, one must allow for the existence of a new mass scale ($M$) and/or timescale ($\Gamma$) that characterise deviations from GR, and these will enter the form of $\{\mt,\gamma\}$.\footnote{The usual quasi-static approximation assumes that $\Gamma\ll k$, allowing such time derivative terms to be neglected; on larger, non-quasistatic scales we must keep them.}

The arguments presented in \cite{Baker:2014tc} motivate the following approximate forms for use \textit{outside} the quasi-static regime:
\begin{align}
\label{jkg}
\mt,\;\gamma&\simeq\frac{{\cal O}\left(1\right)+{\cal O}\left(\frac{\Hu^2}{k^2}\right)+{\cal O}\left(\frac{\Gamma^2}{k^2}\right)+{\cal O}\left(\frac{M^2}{k^2}\right)+\ldots}{{\cal O}\left(1\right)+{\cal O}\left(\frac{\Hu^2}{k^2}\right)+{\cal O}\left(\frac{\Gamma^2}{k^2}\right)+{\cal O}\left(\frac{M^2}{k^2}\right)+\ldots}
\end{align}
It should be understood that all terms are multiplied by time-dependent coefficients specific to a particular gravity theory; in model-independent calculations, a sensible ansatz must be supplied for these. The key point is that $\mt$ and $\gamma$ have the same $k$-dependence as their quasi-static counterparts, with some additional scales $\Gamma$ and $M$ coming into play. Given that the purpose of this paper is to investigate the effects of Hubble-scale corrections to GR, we will take $\Gamma\sim M\sim\Hu$, which further simplifies Eq.~(\ref{jkg}). 

This line of reasoning justifies the use of a simple parameterisation such as Eqs.~(\ref{Poisson_gi}) and (\ref{slip_gi}) to explore large-scale MG phenomenology, which we treat in the next section. An earlier model-independent treatment of horizon scales was incorporated in the Parametrized Post-Friedmann formalism of \cite{PPF2007}, but the parameterisation we use here has the advantages of simplicity and updated knowledge about the generic scale-dependence of MG theories, as discussed above. A scheme for extending beyond the quasi-static regime in Horndeski gravity was very recently presented in \cite{Lombriser2015}, using ideas similar to those discussed above.\footnote{The authors use the evolution at a fixed `pivot scale', $k_*$, to define a timescale equivalent to our $\Gamma$; see Eqs.~(3.1) and (3.2) of \cite{Lombriser2015}.}

In \S\ref{subsec:example} we will evaluate our results numerically, using a simple model based on Eq.~(\ref{jkg}). We do not claim that Eqs.~(\ref{Poisson_gi}--\ref{jkg}) apply comprehensively to all gravity theories on large scales,\footnote{Their derivation relies on assumptions such as that the evolution of a scalar degree of freedom can approximately be described by $\dot\chi\sim\Gamma_\chi \chi$.} only that they are sufficiently general for use in a model-independent investigation such as ours. Similar forms were used (also outside the quasi-static regime) in recent constraints on dark energy by \cite{Planck:2015tt}, for example. Constraints on $\{\mt,\gamma\}$ in two bins of $k$, using data from the WiggleZ, BOSS, and 6dF surveys, were recently presented in \cite{Johnson2015}.

\section{Scale-Dependent Growth}
\label{sec:set-up}

Even with relativistic corrections taken into account, the matter density perturbation $D_{\rm M}$ remains the dominant contribution to the observables considered in \S\ref{sec:CLASS}. In this section, we will build some basic analytic intuition about the effects of our modified equations (\ref{Poisson_gi}) and (\ref{slip_gi}) on the evolution and scale-dependence of $D_{\rm M}$. (We note that the growth rate of matter density perturbations on large scales in a few specific gravity models and the Parametrized Post-Friedmann framework was considered in \citealt{2013PhRvD..87j4019L}).

Issues of gauge choice become crucial when dealing with {\corr non-observable quantities on near-horizon scales}, so we will work with gauge-invariant variables as far as possible. To avoid cluttered expressions, we will suppress the arguments $\left(k,a\right)$ for most variables.

\subsection{Set-up}
\label{subsec:defs}

 Let the line element showing the perturbations of the FLRW metric in a general gauge be \citep{Durrer2005,2011PhRvD..84f3505B}:
\begin{align}
ds^2&=a^2(\eta)\Big\{-(1+2A)d\eta^2-2B_i dx^id\eta \nonumber\\
&+\left[(1+2H_L)\delta_{ij}+2H_{T\,ij}\right]dx^idx^j\Big\}.
\end{align}
{\raggedright As we are interested in only the spin-0 components of the perturbations here, we can write \mbox{$B_i=-k_iB/|k|$} and \mbox{$H_{T\,ij}=(k_ik_j-\frac{k^2}{3}\delta_{ij})H_T/k^2$}.} For a fluid with four-velocity $u^\mu$, we define the spin-0 component of its spatial 3-velocity perturbation to be:
\begin{align}
u^i&=\frac{v^i}{a} & \bf{v}&=-\frac{1}{k}{\bf\nabla}v.
\end{align}

With these definitions in hand, we will work in terms of the following gauge-invariant variables:
\begin{align}
\hat\Psi&=A-\frac{\Hu}{k}\left(\frac{\dot{H}_T}{k}-B\right)-\frac{1}{k}\left(\frac{\ddot{H}_T}{k}-\dot{B}\right)\\
\hat\Phi&=-H_L-\frac{1}{3}H_T+\frac{\Hu}{k}\left(\frac{\dot{H}_T}{k}-B\right)\\
D&=\delta+3(1+w)\frac{\Hu}{k}(v-B)\label{Ddef}\\
V&=v-\frac{\dot{H}_T}{k}\\
Y&=D-3(1+w)\left(\frac{\Hu}{k}V+\hat\Phi\right)\label{Ydef1}\\
&=D+3(1+w)\cR_{\mathrm{co}}\label{Ydef2}.
\end{align}
where the last line above defines $\cR_{\mathrm{co}}$, which becomes equal to (minus) the curvature perturbation in the comoving gauge. 
All of our calculations will concern times well after matter-radiation equality, such that pressureless matter (indicated by a subscript ${\rm M}$) is the dominant clustering matter component and anisotropic stress is negligible.

We need four equations to solve for the four independent variables $\hat\Phi$, $\hat\Psi$, $D_{\rm M}$ and $V_{\rm M}$ (from Eq.~(\ref{Ydef1}) we see that $Y_{\rm M}$ is simply a linear combination of these). Two are provided by Eqs.~(\ref{Poisson_gi}) and (\ref{slip_gi}), where it should be noted that $\{\mt,\gamma\}$ are defined in a gauge-independent manner. The other two are the standard fluid equations for pressureless matter perturbations:
\begin{align}
\dot{Y}_{\rm M}&=-kV_{\rm M} & \dot{V}_{\rm M}&=-\Hu V_{\rm M}+k\hat\Psi \label{consv},
\end{align}
where overdots denote conformal time derivatives. The usual ${\delta}_{\rm M}$ term is sitting inside our variable ${Y}_{\rm M}$. {\corr Note that we have assumed here that pressureless matter follows its standard conservation laws, thereby excluding some coupled dark energy models (for example) from our treatment. 
We leave such extensions to a future work; note that there will be some resulting changes to Eq.~(\ref{eq:DeltaN}).}

\subsection{Evolution of $Y_{\rm M}$ and $D_{\rm M}$}

The `cleanest' variable to solve for is $Y_{\rm M}$. If we solve for $Y_{\rm M}$ instead of $D_{\rm M}$ directly, we avoid the need to take time derivatives of our parameterisation functions $\mt$ and $\gamma$; this is advantageous both for simplicity's sake, and also because $\mt$ and $\gamma$ are effective, rather than exact, objects.

Substituting Eqs.~(\ref{slip_gi}), (\ref{Ydef1}) and the second of Eqs.~(\ref{consv}) into Eq.~(\ref{Poisson_gi}), and using $3\Hu^2\Omega_{\rm M}=8\pi G_N a^2\rho_{\rm M}$, we obtain
\begin{align}
\label{PhiYrel}
-2k^2\hat\Psi&=3\Hu^2\Omega_{\rm M}\mt\frac{\left(Y_{\rm M}-3\frac{\Hu}{k^2}\dot{Y}_{\rm M}\right)}{\left(1+\frac{9}{2}\frac{\Hu^2}{k^2}\Omega_{\rm M}\mt\gamma\right)}.
\end{align}
Note that the matter fraction here is time-dependent; we will use $\Omega_{{\rm M}0}$ for its value today.
 Comparing Eqs.~(\ref{PhiYrel}) and (\ref{Poisson_gi}), we infer that 
 \begin{align}
\label{DelYrel}
D_{\rm M}&=\frac{\left(Y_{\rm M}-3\frac{\Hu}{k^2}\dot{Y}_{\rm M}\right)}{\left(1+\frac{9}{2}\frac{\Hu^2}{k^2}\Omega_{\rm M}\mt\gamma\right)}.
\end{align}
Differentiating the first of Eqs. (\ref{consv}) and using Eqs.~(\ref{slip_gi}), (\ref{PhiYrel}) and the second of (\ref{consv}) then leads us to an evolution equation for $Y_{\rm M}$,
\begin{align}
\label{ddY}
{\ddot Y}_{\rm M}&+\Hu {\dot Y}_{\rm M}\left[1+\frac{\A\mt}{1+\A \mt\gamma}\right]\nonumber\\
&-\frac{3}{2}\Hu^2\Omega_{\rm M}\,\mt\,Y_{\rm M}\left[\frac{1}{1+\A\mt\gamma}\right]=0.
\end{align}
The solutions of this equation (as a function of $k$ and $z$ or $a$) can then be used to find the evolution of the matter density perturbation, $D_{\rm M}$, using Eq.~(\ref{DelYrel}).

\subsection{Sub- and Super-Horizon Limits}
\label{subsec:limits}

Using the parameterised Poisson equation, Eq.~(\ref{Ydef1}) can be rewritten as
\begin{align}
Y_{\rm M}&=D_{\rm M}\left(1+\frac{9\Hu^2\Omega_{\rm M}\mt\gamma}{2k^2}\right)-3\frac{\Hu}{k}V_{\rm M}.
\end{align}
In the subhorizon limit, $k\gg\Hu$, this gives $Y_{\rm M}\rightarrow D_{\rm M}$,\footnote{Assuming that $\mt\gamma$ becomes a scale-independent function of time in this limit, which is true of Eqs.~(\ref{jkg}),  (\ref{mu_ansatz}) and (\ref{gamma_ansatz}).} which is a manifestation of the fact that gauge differences are generally unimportant on small distance scales. In this limit, Eq.~(\ref{ddY}) recovers the usual evolution equation for $D_{\rm M}$ on small scales (e.g. \citealt{Amendola_book,Baker:2013hia}),
\begin{align}
\label{QSY}
{\ddot Y}_{\rm M}+\Hu {\dot Y}_{\rm M}-\frac{3}{2}\Hu^2\Omega_{\rm M}\,\mt\,Y_{\rm M}=0 .
\end{align} 
We see that small scales are sensitive \textit{only} to $\mt$ and not $\gamma$, with enhanced growth (relative to GR) expected for values $\mt>1$. 

\begin{figure}[t]
\hspace{-1.4em}\includegraphics[width=1.05\columnwidth]{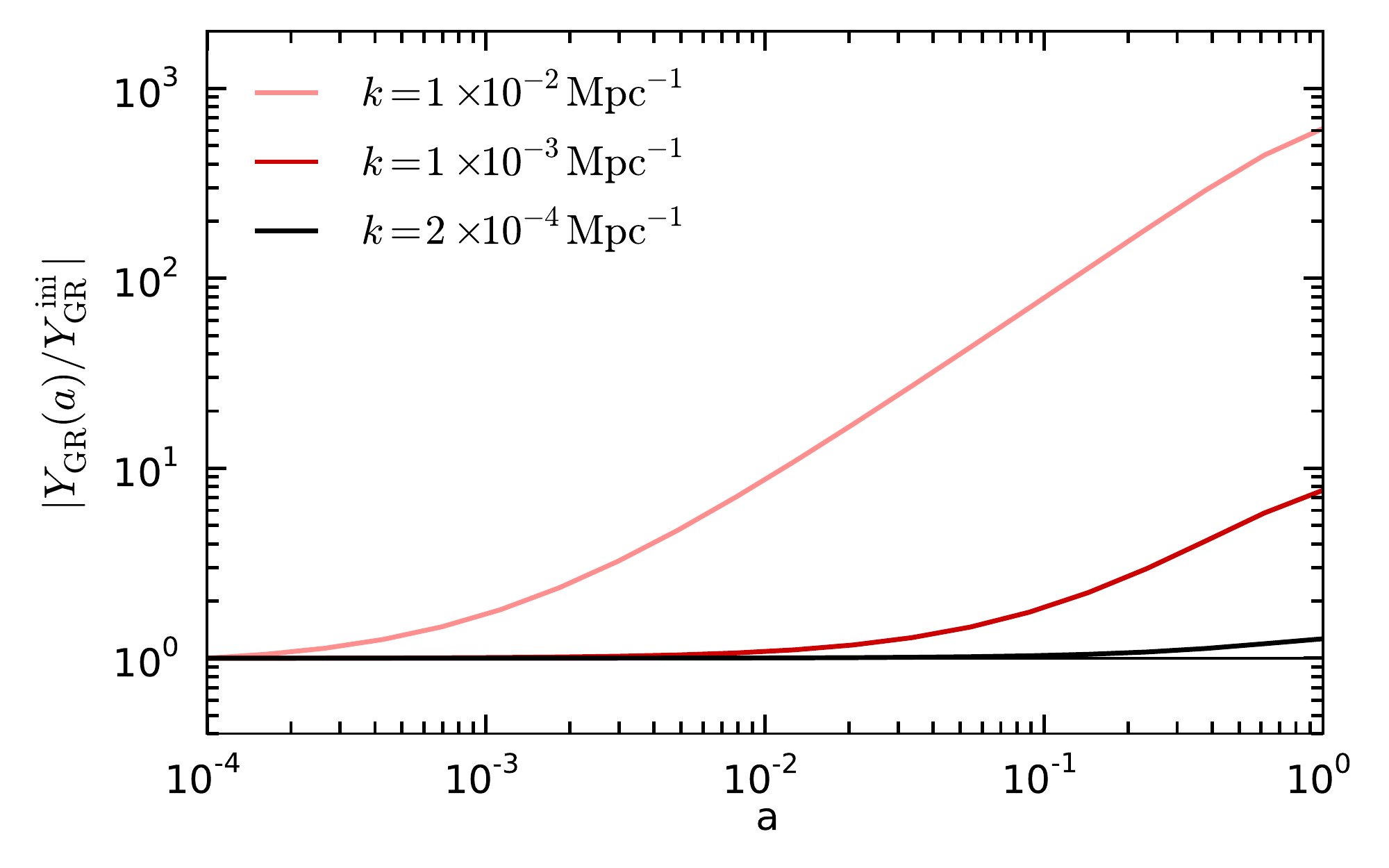}
\vspace{-0.4cm}\caption{Growth of the variable $Y_{\rm M}$ (defined in Eq.~\ref{Ydef1}) in GR, for different wavemodes. Note that the large-scale mode $k=2\times 10^{-4}$ Mpc$^{-1}$ scarcely evolves. Curves have been normalised to their initial values, which depends on $k$.}
\label{fig:Ygr}
\end{figure}

To understand the behaviour of superhorizon scales, let us first consider the GR case. In the limit $k\ll\Hu$ and $\mt=\gamma=1$, Eq.~(\ref{ddY}) becomes
\begin{align}
\ddot{Y}_{\rm M}^{\rm GR}+2\Hu\dot{Y}_{\rm M}^{\rm GR}\simeq 0,
\label{YULSGR}
\end{align}
that is, the equation for $Y_{\rm M}$ becomes unsourced and hence is solved by $Y_{\rm M}^{\rm GR}\rightarrow$ constant on ultra-large scales; see Fig.~\ref{fig:Ygr}. (The second solution for the above equation blows up as $a\rightarrow 0$, and so must be removed by choice of boundary conditions.)

To proceed, we now consider small deviations of quantities from their GR values, writing \mbox{$Y_{\rm M}=Y_{\rm M}^{\rm GR}+\delta Y_{\rm M}$}, \mbox{$D_{\rm M}=D_{\rm M}^{\rm GR}+\delta D_{\rm M}$}, \mbox{$\mt=1+\delta\mt$} and $\gamma=1+\delta\gamma$. For brevity we will not display the full linearised versions of Eqs.~(\ref{DelYrel}) and (\ref{ddY}), but their large-scale limits are
\begin{align}
\delta \ddot{Y}_{\rm M}&+2\Hu \delta \dot{Y}_{\rm M} -\Hu \dot{Y}_{\rm GR}\delta\gamma\simeq 0 \label{deltaY}\\
\frac{\delta D_{\rm M}}{D_{\rm M}^{\rm GR}}&\simeq \frac{\delta \dot{Y}_{\rm M}}{\dot{Y}_{\rm M}^{\rm GR}}-\left(\delta\tilde\mu+\delta\gamma\right)\label{deltadeltaM}.
\end{align}
Note that although $\dot{Y}_{\rm M}^{\rm GR}\rightarrow 0$ for $k\rightarrow 0$, the first term in Eq.~(\ref{deltadeltaM}) does not have to diverge if $\delta {\dot Y}_{\rm M}$ is decreasing with $k$ at a similar rate. In Appendix~\ref{app:DM}, we show that this ratio is given on ultra-large scales by
\begin{align}
\label{Yint}
\frac{\delta \dot{Y}_{\rm M}}{\dot{Y}_{\rm M}^{\rm GR}}\simeq \int\left(\Hu \,\delta\gamma\right)d\eta.
\end{align}
From Eq.~(\ref{deltadeltaM}) we see that the large-scale behaviour of the dark matter density perturbation $D_{\rm M}$ is controlled by the interplay of two terms. For a late-time modification of gravity, one expects the integral in Eq.~(\ref{Yint}) to be the subdominant term, as the integrand is suppressed by the decreasing Hubble factor just when $\delta\gamma$ begins to grow. These formulae will help us to understand the large-scale behaviour of the two numerical examples presented in the next subsection. 

Note that whilst small scales were only sensitive to $\mt$, large scales depend on both $\mt$ and $\gamma$ via Eq.~(\ref{deltadeltaM}). Also note that the influence of $\mt$ changes directionality: on small scales, $\mt >1$ enhances the growth of CDM perturbations, whilst on ultra-large scales it suppresses growth relative to the GR case \citep[this feature was also noticed in][]{2013PhRvD..87f4026H}.

In between the the large- and small- scale limits exists a transition regime, requiring the full numerical solution of Eq.~(\ref{ddY}). It is this regime which is the most interesting observationally -- even the next generation of surveys will not probe scales large enough to see the $k\ll\Hu$
limit at late times (corresponding to the area left of the solid blue line in Fig.~\ref{fig:horizon}). In the next subsection we solve for the full range of $k$ for two contrasting examples of $\{\mt, \gamma\}$. 

\begin{figure*}[t]
  \begin{minipage}[b]{0.52\linewidth}
    \centering
    \hspace{-3em}\includegraphics[width=\columnwidth]{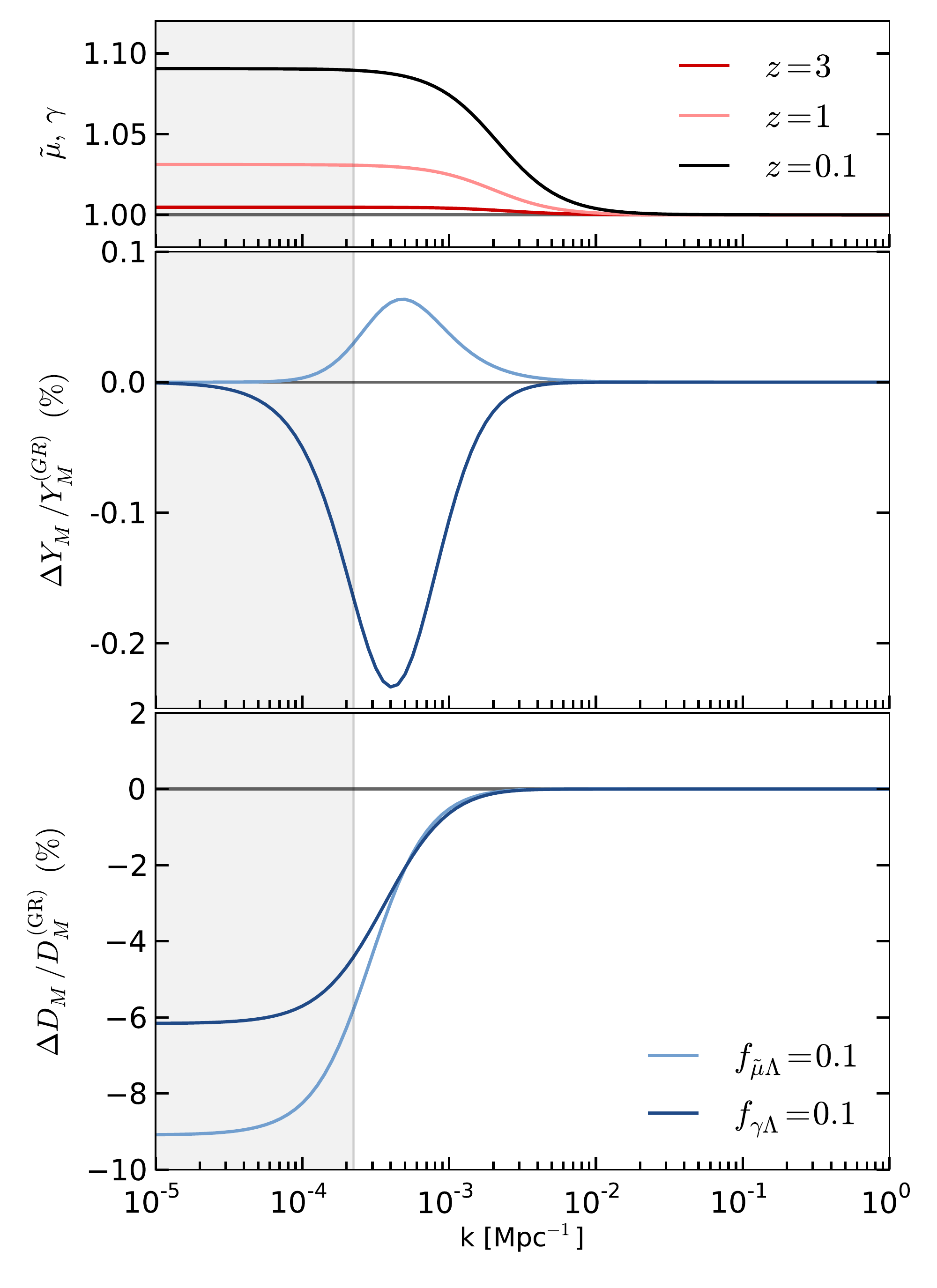}
  \end{minipage}%
  \begin{minipage}[b]{0.52\linewidth}
    \centering
    \hspace{-3em}\includegraphics[width=\columnwidth]{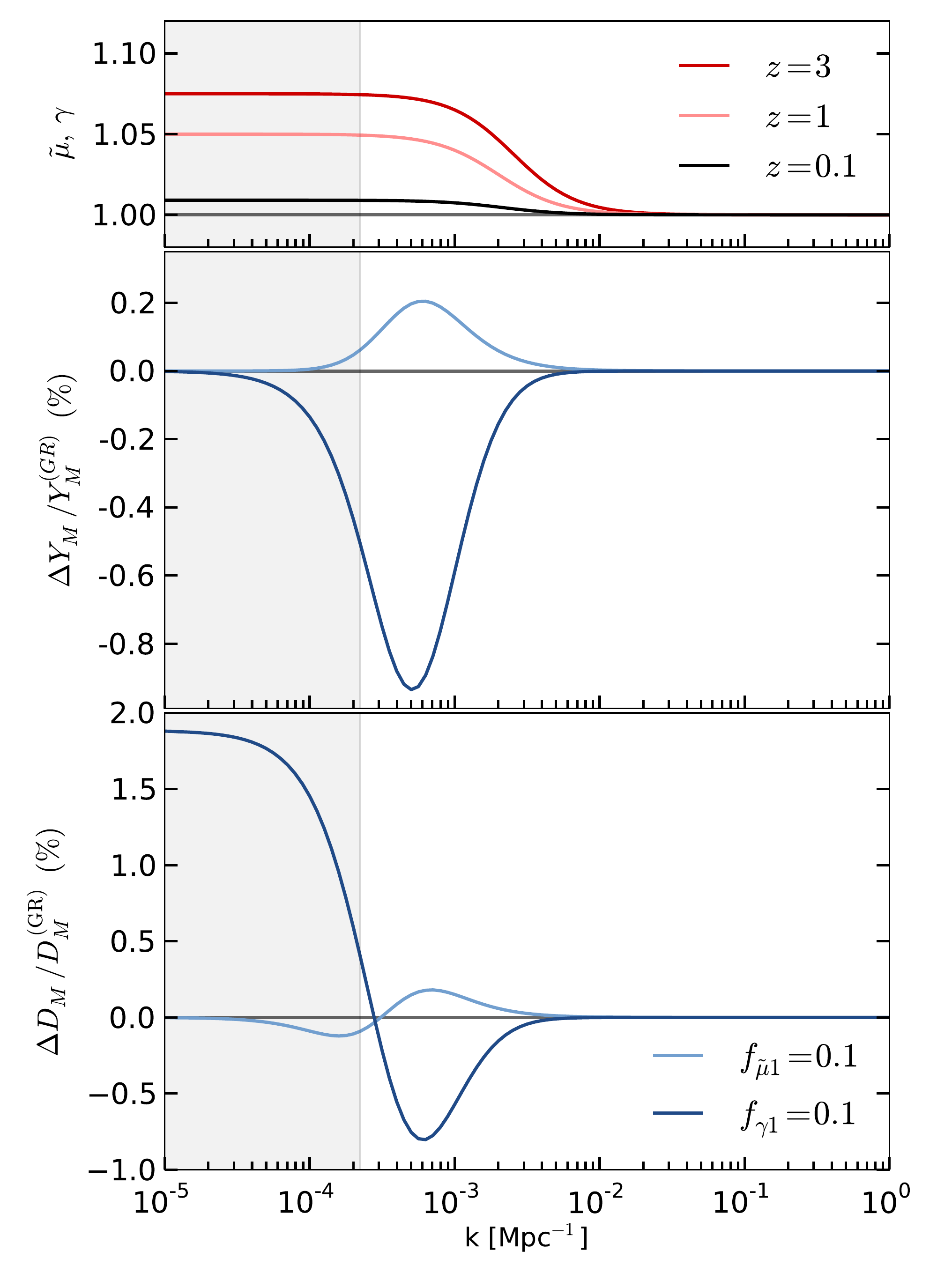}
  \end{minipage}
\caption{Growth of $Y_{\rm M}$ and $D_{\rm M}$, as defined in Eqs.~(\ref{Ddef}) and (\ref{Ydef1}). The left-hand panel employs the late-time emergence ($\Lambda$-type) ansatz, the right-hand panel uses the early-time ansatz (`CPL-like'), see Eqs.~(\ref{OmL_ansatz}-\ref{CPL_ansatz2}). The dark and pale curves in the two lower panels correspond to cases with $\delta\mt$ and $\delta\gamma$ switched on individually, respectively. The top panel shows the time- and scale- dependence of either $\mt$ or $\gamma$, which control the deviations from GR. Note that although in the CPL-like case $\mt(z=0)=1$, $\gamma(z=0)=1$, there is an integrated effect from the modified field equations that persists to late times.}
\label{fig:OmL_DM}\label{fig:CPL_DM}
\end{figure*}

\subsection{Example Ansatz}
\label{subsec:example}

The derivations of the previous section are independent of the precise forms of the parameterisation functions $\mt$ and $\gamma$. To solve for the full near-horizon behaviour, we now need to pick concrete forms for $\mt$ and $\gamma$.

As argued in \S\ref{sec:params} and \cite{Baker:2014tc}, a motivated scale-dependent form of $\mt$ and $\gamma$ (for theories with one new d.o.f.) is a ratio of polynomials in $(M/k)^2$ and $(\Gamma/k)^2$, where $M$ and $\Gamma$ are a mass scale and timescale characteristic of any modifications to GR. In this paper we wish to explore scenarios that preserve GR at short distances whilst allowing new phenomena to emerge in the regime $k\sim\Hu$. For the present case, then, we simply take $M\sim\Gamma\sim\Hu$.

The precise form of Eq.~(\ref{jkg}) is difficult to explore in a model-independent manner, as one then has no guidance for how the time-dependent coefficients are related to one another. We will therefore use a slightly altered version of this scale-dependence that makes preservation of the GR limit more transparent \citep[a similar form was used in Eqs.~(46) and (47) of][]{Planck:2015tt}:
\begin{align}
\mt(a,k)&=1+f_\mt(a)\left[\frac{\left(\frac{\lambda\Hu}{k}\right)^2}{1+\left(\frac{\lambda\Hu}{k}\right)^2}\right] \label{mu_ansatz}\\
\gamma(a,k)&=1+f_\gamma(a)\left[\frac{\left(\frac{\lambda\Hu}{k}\right)^2}{1+\left(\frac{\lambda\Hu}{k}\right)^2}\right] \label{gamma_ansatz}.
\end{align}
Here, $f_\mt(a)$ and $f_\gamma(a)$ are functions of scale factor that control the time evolution of MG effects, and $\lambda$ is a parameter that determines the location (in Fourier space) of the transition from the small-scale GR regime to the large-scale MG regime. We will adopt $\lambda=10$ in this work, as this allows MG effects to become important only on scales larger than $k_{\rm eq}$, which is effectively the largest scale probed by large-scale structure surveys to date. While picking $\lambda = 1$ may seem more natural, this would limit deviations from GR to multipoles $\ell \lesssim 3$, as shown in Fig.~\ref{fig:horizon}. (We note in passing that subhorizon transition scales can occur in models of a clustering dark energy component with subluminal sound speed.)

We must also choose some parameterisations of the time-dependence of $f_\mt(a)$ and $f_\gamma(a)$ to explore. We compare two ansatzes that have opposite behaviours; one where the MG functions grow larger with time, and another in which they decay. This will allow us to explore the interplay between different terms on ultra-large scales found in Eq.~(\ref{deltadeltaM}). To facilitate comparison with the existing literature, we used the two time-dependent ansatzes considered by \cite{Planck:2015tt} for their constraints on dark energy and gravity:
\begin{enumerate}
\item $\Lambda$-like/late-time emergence (recall that we invoke a $\Lambda$CDM expansion history):
\begin{align}
 f_\mt(a)&=f_{\mt\Lambda}\,\frac{\Omega_{\Lambda}(a)}{\Omega_{\Lambda 0}}, &  f_\gamma(a)&=f_{\gamma\Lambda}\,\frac{\Omega_{\Lambda}(a)}{\Omega_{\Lambda 0}} 
 \label{OmL_ansatz}
 \end{align}
\item Behaviour analogous to the Chevallier-Polarski-Linder \citep[CPL;][]{Chevallier2001, Linder2003} equation of state parameterisation:
\begin{align}
 f_{\mt}(a)&=f_{\mt 0}+f_{\mt 1}(1-a)  \label{CPL_ansatz1}
 \\
 f_{\gamma}(a)&=f_{\gamma 0}+f_{\gamma 1}(1-a)
  \label{CPL_ansatz2}.
 \end{align}
\end{enumerate}
Option (1) is the natural choice for MG effects that hope to explain late-time acceleration, whilst option (2) hypothesises that any deviation from an equation of state of $w=-1$ might be accompanied by modified perturbations. Given the tight constraints on early dark energy from the CMB \citep{2011PhRvD..83l3504C}, we impose an additional cutoff in our numerical implementation that smoothly brings the modifications to zero at $z\gtrsim 5$.
 
Fig.~\ref{fig:OmL_DM} shows the deviations of $Y_{\rm M}$ and $D_{\rm M}$ as ratios of their GR values for both ansatzes, as a function of $k$ at $z=0$. For the $\Lambda$-like ansatz (left panel), two cases are plotted: $\{f_{\mt\Lambda},\,f_{\gamma\Lambda}\}=\{0.1,0\},\;\{0,0.1\}$. We have chosen to vary $\mt$ and $\gamma$ individually in order show their distinct effects; in a realistic MG model one would typically expect them to vary simultaneously. 
The 10\% amplitude of deviations from GR was chosen simply for illustration.

The right panel side of Fig.~\ref{fig:OmL_DM} shows analogous plots for the CPL-like ansatz. The curves plotted correspond to $\{f_{\mt 1},\,f_{\gamma 1}\}=\{0.1,0\},\;\{0,0.1\}$, with $\{f_{\mt 0},\,f_{\gamma 0}\}=\{0,0\}$ for both; once again, this particular example is motivated only by simplicity.

 The top row in both panels shows the time- and scale-dependent form adopted by either $\mt$ or $\gamma$. Note that the transition away from the GR regime extends to wave numbers some way below the horizon. The following other features are also of interest:
 \begin{itemize}
\item The ultra large-scale behaviour of the matter density perturbation can be understood from Eqs.~(\ref{deltadeltaM}) and (\ref{Yint}). In Fig.~\ref{fig:OmL_DM}, where the MG effects become relevant at late times, the integral term in Eq.~(\ref{deltadeltaM}) is subdominant, so the large-scale value of $D_{\rm M}/D_{\rm M}^{\rm GR}$ is primarily set by the maximum of $\delta\mt$ or $\delta\gamma$. (The offset between the curves is due to the integral contribution to $\gamma$.)
\item In the CPL-like case, the integrand of Eq.~(\ref{Yint}) is larger, and this dominates the large-scale deviation if $\delta\gamma\neq 0$. Since $\delta\mt\rightarrow 0$ as $z\rightarrow 0$, the deviations vanish on large scales for $\{f_{\mt 1},\,f_{\gamma 1}\}=\{0.1,0\}$. 
\item As mentioned in \S\ref{subsec:limits}, horizon-scale modifications of the effective gravitational constant suppress growth; this contrasts with the enhanced growth they cause on subhorizon scales. 
\item The time-dependence of MG effects plays a strong role in their impact on the density perturbations; compare the $\sim10\%$ deviations for the $\Lambda$-like ansatz with only $\sim 2\%$ deviations for the CPL one.
\end{itemize}
Much of this (qualitative) behaviour is mirrored in the observable angular power spectra calculated in \S\ref{sec:CLASS}, as the density-density (and density-RSD) term is dominant over a wide range of scales and redshifts \citep{2011PhRvD..84f3505B}.

\subsection{Comoving Curvature Perturbation} 
\label{subsec:comoving}

Starting from Eqs.~(\ref{Poisson_gi}), (\ref{slip_gi}) and (\ref{consv}), one can show that the comoving curvature perturbation \mbox{$\cR_{\mathrm{co}}=-\hat\Phi-V \Hu/k$} is conserved on ultra-large scales, provided that $\mt$ does not tend to zero there. We relegate the derivation to Appendix~\ref{app:Rconsv}. This is consistent with the well-known results of \cite{Wands:2000jd} and \cite{Bertschinger:2006ht}, who showed that this feature should hold true in any metric theory of gravity. 

The conservation of \mbox{$\cR_{\mathrm{co}}$}, together with the conservation equations for CDM, immediately imply that the evolution of the metric potentials on superhorizon scales is given by
\begin{align}
\ddot{\hat\Phi}+\Hu\dot{\hat\Psi}-\dot{\hat\Phi}\left(\frac{3\Hu\dot\Hu-\ddot\Hu-\Hu^3}{\Hu^2-\dot\Hu}\right)+\hat\Psi\frac{(\Hu\ddot\Hu-2{\dot\Hu}^2)}{(\Hu^2-\dot\Hu)}=0.
\label{ddPhiULS}
\end{align}
As noted in \cite{Bertschinger:2006ht} and \cite{2013PhRvD..87f4026H}, this equation completely fixes the evolution of superhorizon potentials in terms of the cosmological expansion rate once a relationship between $\hat\Phi$ and $\hat\Psi$ has been specified. If $\gamma\neq 1$, this means that deviations from GR are expected on superhorizon scales, even if the background expansion history is identical to that of $\Lambda$CDM. 

Notice that it is the slip ratio, $\gamma$, that is most relevant to superhorizon behaviour (see also Eq.~\ref{deltaY}), in contrast to the controlling influence of $\mt$ on small scales (Eq.~\ref{QSY}).

\begin{figure*}[t]
  \begin{minipage}[b]{0.5\linewidth}
    \centering
    \hspace{-2em}\includegraphics[width=\columnwidth]{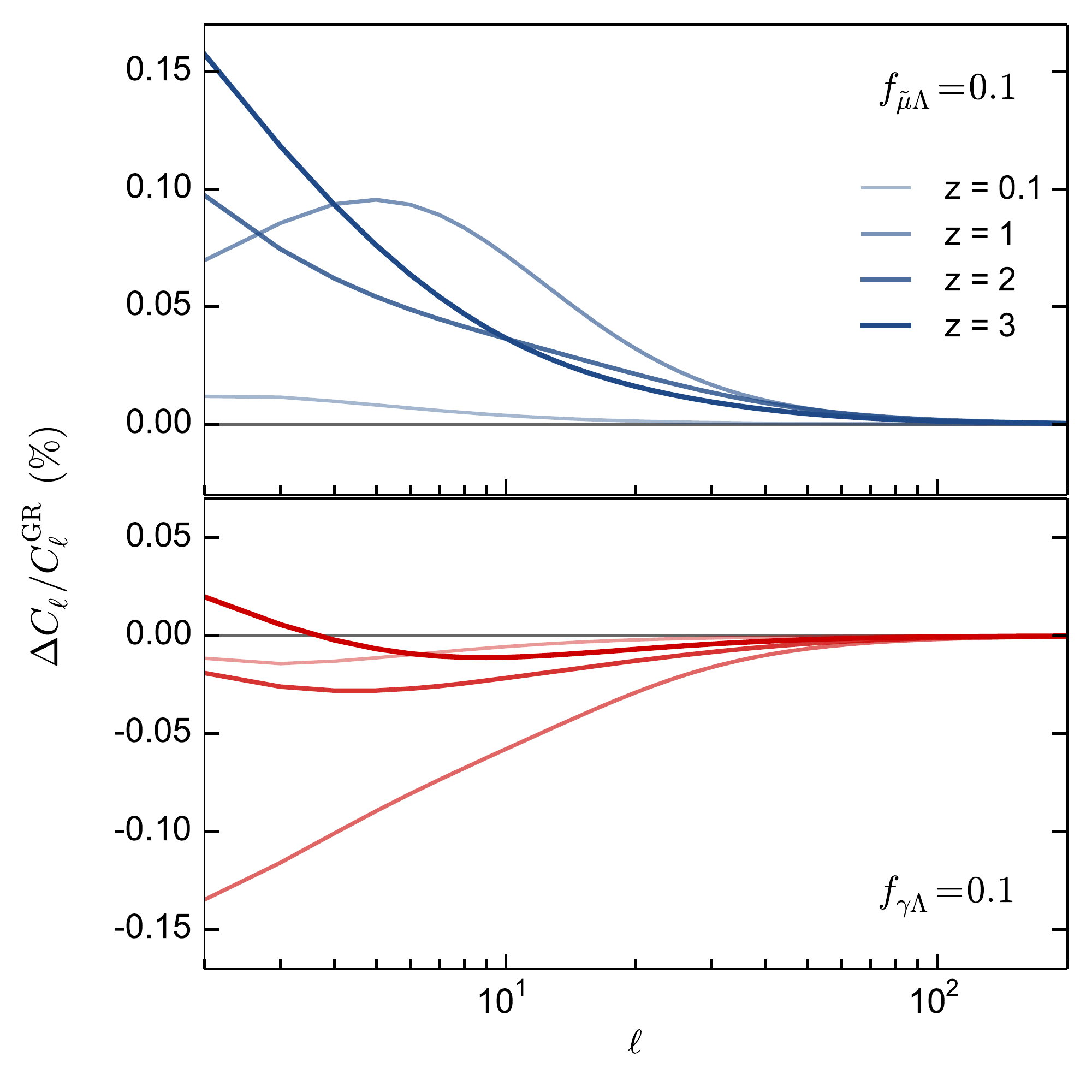}
  \end{minipage}%
  \begin{minipage}[b]{0.5\linewidth}
    \centering
    \hspace{-2em}\includegraphics[width=\columnwidth]{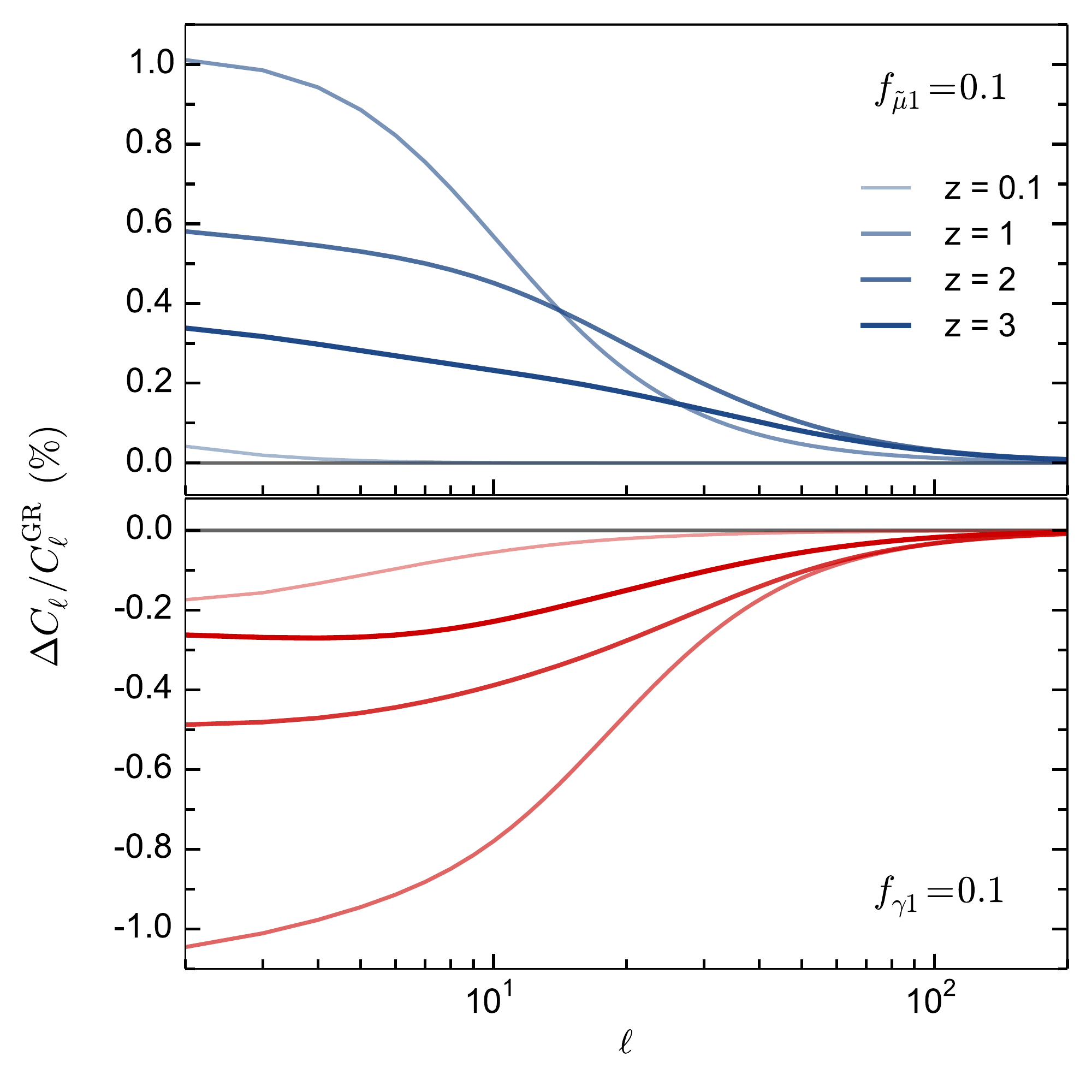}
  \end{minipage}
\caption{Fractional deviation of the number count power spectrum from its GR behaviour, for the $\Lambda$ (left) and CPL-like (right) parameterisations. Four redshift bins are plotted for each set of parameters, all with width $\Delta z = 0.1$. We have fixed $\lambda = 10$.}\vspace{0.8em}
\label{fig:deltacl}
\end{figure*}

\section{Ultra-large scale observables} 
\label{sec:CLASS}

Observables such as galaxy number counts are primarily sensitive to fluctuations in matter density, which we explored in \S\ref{sec:set-up}, but a number of other effects can also contribute corrections to the observed quantity. Lensing by intervening matter, peculiar velocities of the sources, evolution of potentials along the line of sight, and several purely relativistic effects lead to distortions of the redshift-space positions of objects, their flux distribution, and the survey volume. All of these effects must be taken into account in a fully correct treatment -- especially on ultra-large scales, where the corrections can be large. Complete calculations up to first and even second order in perturbations have been performed for several observables \citep{Bonvin:2005ps, 2011PhRvD..84f3505B, 2011PhRvD..84d3516C, Jeong:2011as, 2012PhRvD..86f3514Y, Umeh:2012pn, 2013PhRvD..87f4026H, Andrianomena:2014sya}, some of which have been implemented in the CLASS and CAMB codes \citep{2011PhRvD..84d3516C, DiDio:2013bqa}.

Previous work on tests of GR on ultra large-scales has mostly focused on the integrated Sachs-Wolfe (ISW) effect \cite[e.g.][]{PhysRevD.69.044005, 2006PhRvD..73l3504Z, 2007PhRvD..75d4004S}, which is a sub-dominant contribution to the relativistic observables. More extensive treatments have been given by \cite{2012PhRvD..86f3514Y, 2013PhRvD..87j4019L} and \cite{2013PhRvD..87f4026H}, who calculate the full linear relativistic expressions for perturbed source number counts and 21cm brightness temperature fluctuations respectively. The former propose to test GR by separately measuring the relativistic `correction' terms that are typically neglected on subhorizon scales, and calculate these effects for several different MG theories in \cite{2013PhRvD..87j4019L}. \cite{2013PhRvD..87f4026H} specialise to $f(R)$ gravity, and perform a principal component analysis of the 21cm power spectrum (although they do not specifically separate out large-scale modes).

We will take a different approach, focusing exclusively on Hubble-scale corrections to GR using the model-independent ansatzes of Eqs.~(\ref{OmL_ansatz}-\ref{CPL_ansatz2}). The observable that we consider is the total linear perturbation to the number counts of sources in a flux-limited redshift survey \citep{DiDio:2013bqa},
\bea
\Delta_{\rm N}(\mathbf{\hat{n}}, k, z) &=& b D_{\rm M} + {\corr (f^{\rm N}_{\rm evo}\! -\! 3)\frac{\aH}{k}V_{\rm M}} -\!\frac{k}{\aH} (\mathbf{\hat{k}}\cdot\mathbf{\hat{n}})^2 V_{\rm M} \label{eq:DeltaN} \\
&+& \Psi + (5s - 2)\Phi + \aH^{-1} \dot{\Phi} \nonumber\\
&+& \frac{2 - 5s}{2 r_S} \int_0^{r_S} dr \left [ 2 - \frac{r_S - r}{r} \nabla_\Omega\right] \left ( \Phi + \Psi \right ) \nonumber \\
&+& \left ( \frac{\dot\aH}{\aH^2} + \frac{2 - 5s}{r_S \aH} + 5s - f^{\rm N}_{\rm evo} \right ) \nonumber \\
&& ~~\times \left ( \Psi + i (\mathbf{\hat{k}}\cdot\mathbf{\hat{n}}) V_{\rm M} + \int_0^{r_S} dr\, (\dot\Phi + \dot\Psi) \right ). \nonumber
\eea
Here, $b$ is the linear bias, $s$ is the magnification bias (the logarithmic derivative of the source number density with respect to limiting magnitude), $f^{\rm N}_{\rm evo}$ is the evolution bias (which describes the proper evolution of the source number density), and $\nabla_\Omega$ is the angular Laplacian. Subscript $S$ denotes evaluation at the source. The expression for the fractional perturbation to the 21cm brightness temperature can be obtained directly from this expression simply by setting $s = 2/5$ \citep{2013PhRvD..87f4026H}. Also see \cite{2012PhRvD..85d1301B} for a discussion of how to define the bias on ultra-large scales.

Many of the terms in Eq.~(\ref{eq:DeltaN}) are well known, and some have already been measured by large-scale structure surveys. The first line consists of the source density (including a gauge correction) and a redshift-space distortion term ($\propto \partial_r(\mathbf{V}\cdot\mathbf{\hat{n}})$). The whole third line is the lensing convergence, and the final term on the last line is the integrated Sachs-Wolfe (ISW) effect, $\propto (\dot{\Phi} + \dot{\Psi})$.

Most of the other terms are strongly suppressed in the subhorizon, quasi-static regime, as they scale as $\aH/k$ or $(\aH/k)^2$ with respect to the dominant density term. As such, they can be neglected to a very good approximation by essentially all large-scale structure surveys to date. As discussed previously, this will no longer be the case with future surveys however, which will be able to access considerably larger scales.

Each of the relativistic correction terms will be modified in different ways by the introduction of modifications to GR, and so the sum of these terms could potentially exhibit interesting features in a MG theory. The question of how the total observable is modified, and whether those modifications can be measured, is the focus of the rest of this section.

\begin{figure*}[t]
\hspace{-1.1em}\includegraphics[height=1.37\columnwidth]{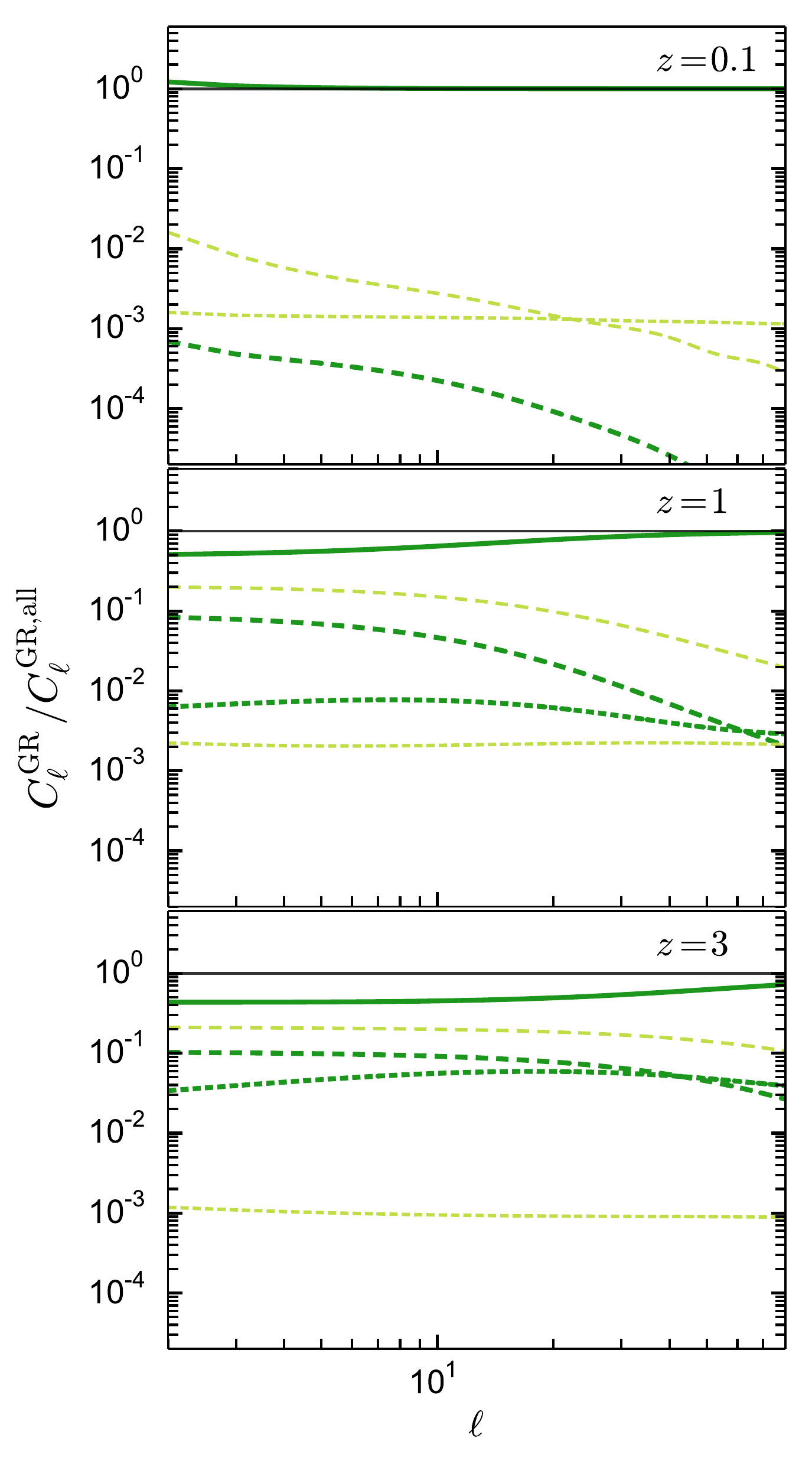}
\includegraphics[height=1.37\columnwidth]{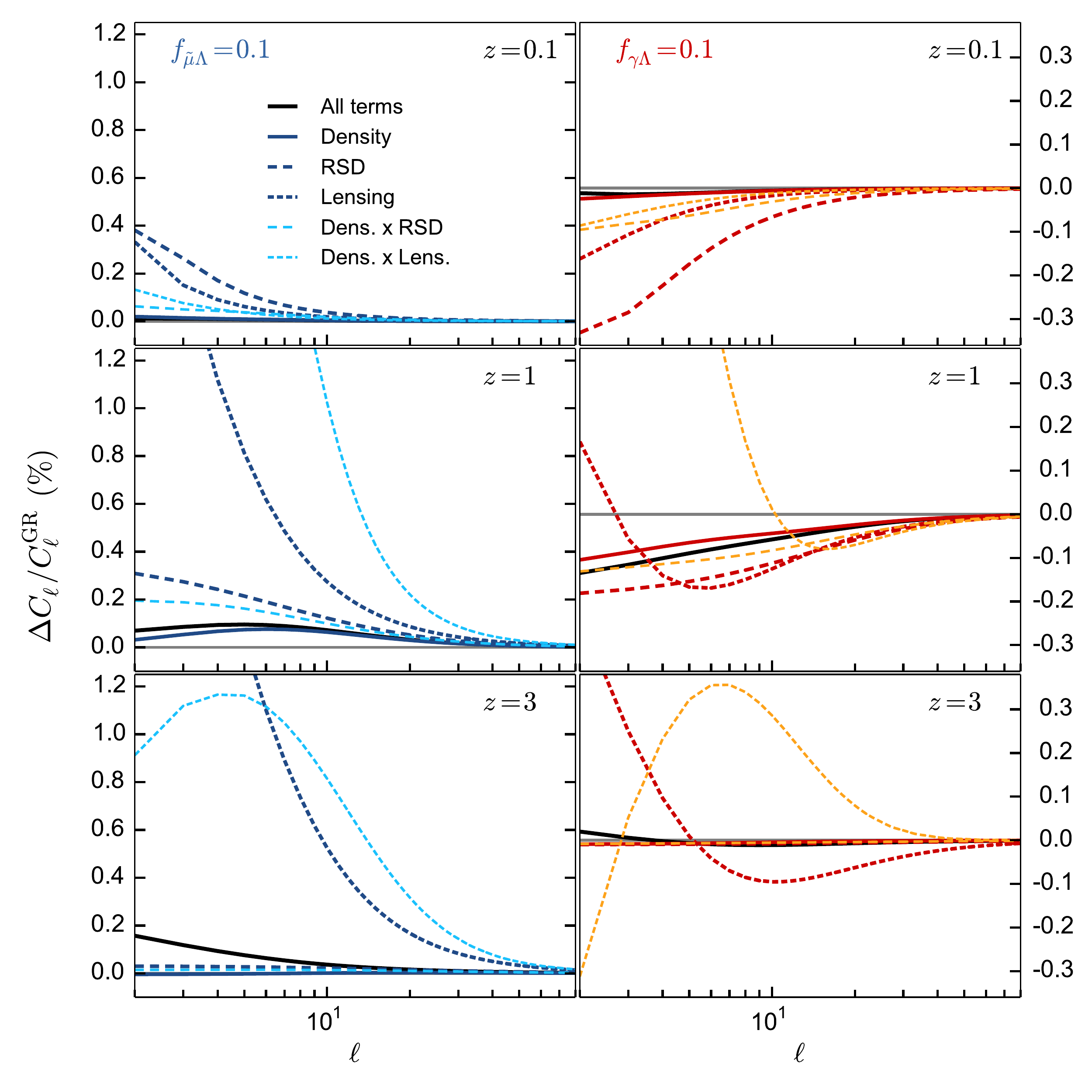}
\caption{{\corr Left panel: Angular power spectra for individual terms from Eq.~(\ref{eq:DeltaN}), for GR, relative to the total number count power spectrum (see right panels for key). Right panels: Fractional deviation of the number count power spectrum for the $\Lambda$ parameterisation, as in Fig.~\ref{fig:deltacl}, for the same terms from Eq.~(\ref{eq:DeltaN}).}}
\label{fig:deltacl_terms}
\end{figure*}

\begin{figure*}[t]
  \begin{minipage}[b]{0.5\linewidth}
    \centering
    \hspace{-2em}\includegraphics[width=\columnwidth]{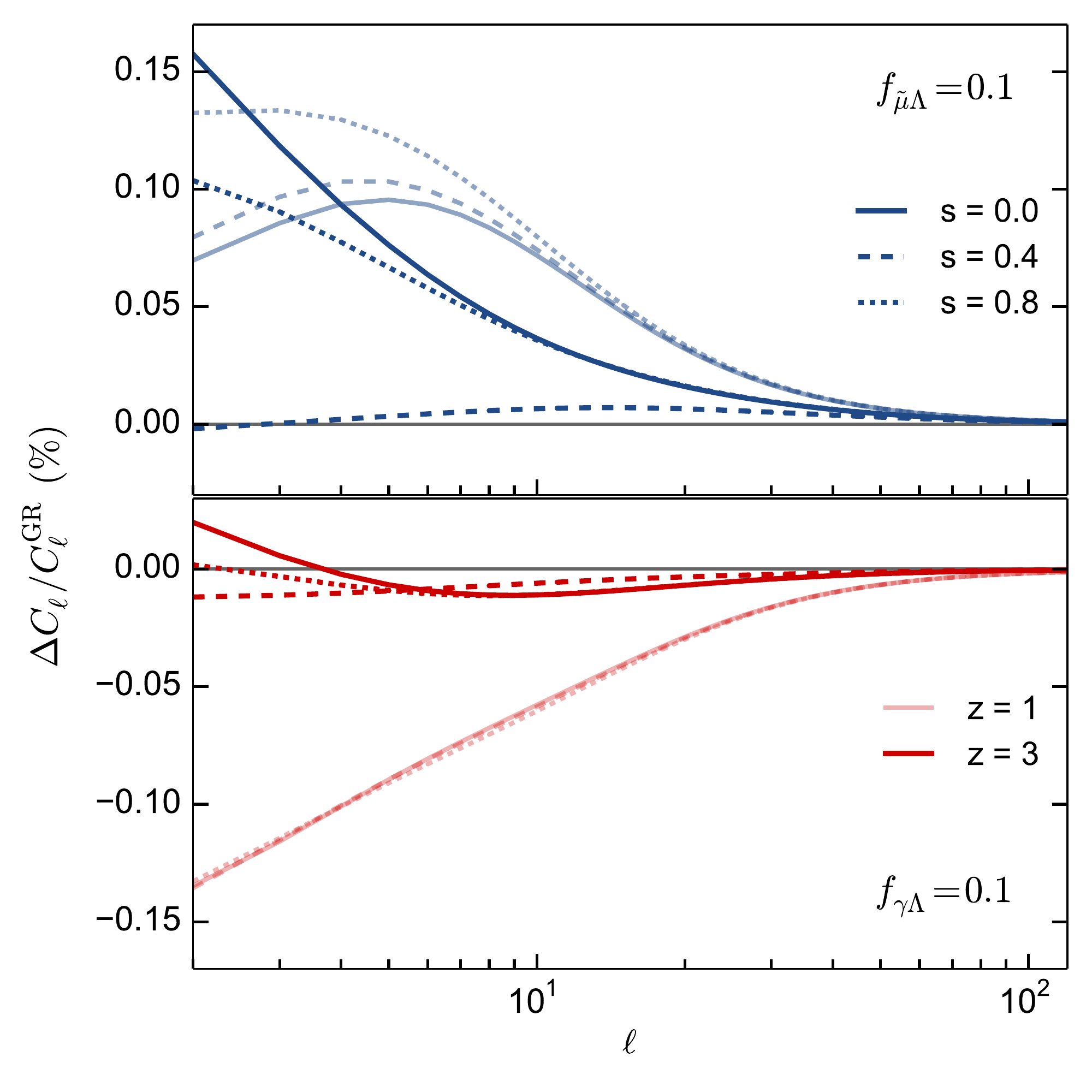}
  \end{minipage}%
  \begin{minipage}[b]{0.5\linewidth}
    \centering
    \hspace{-2em}\includegraphics[width=\columnwidth]{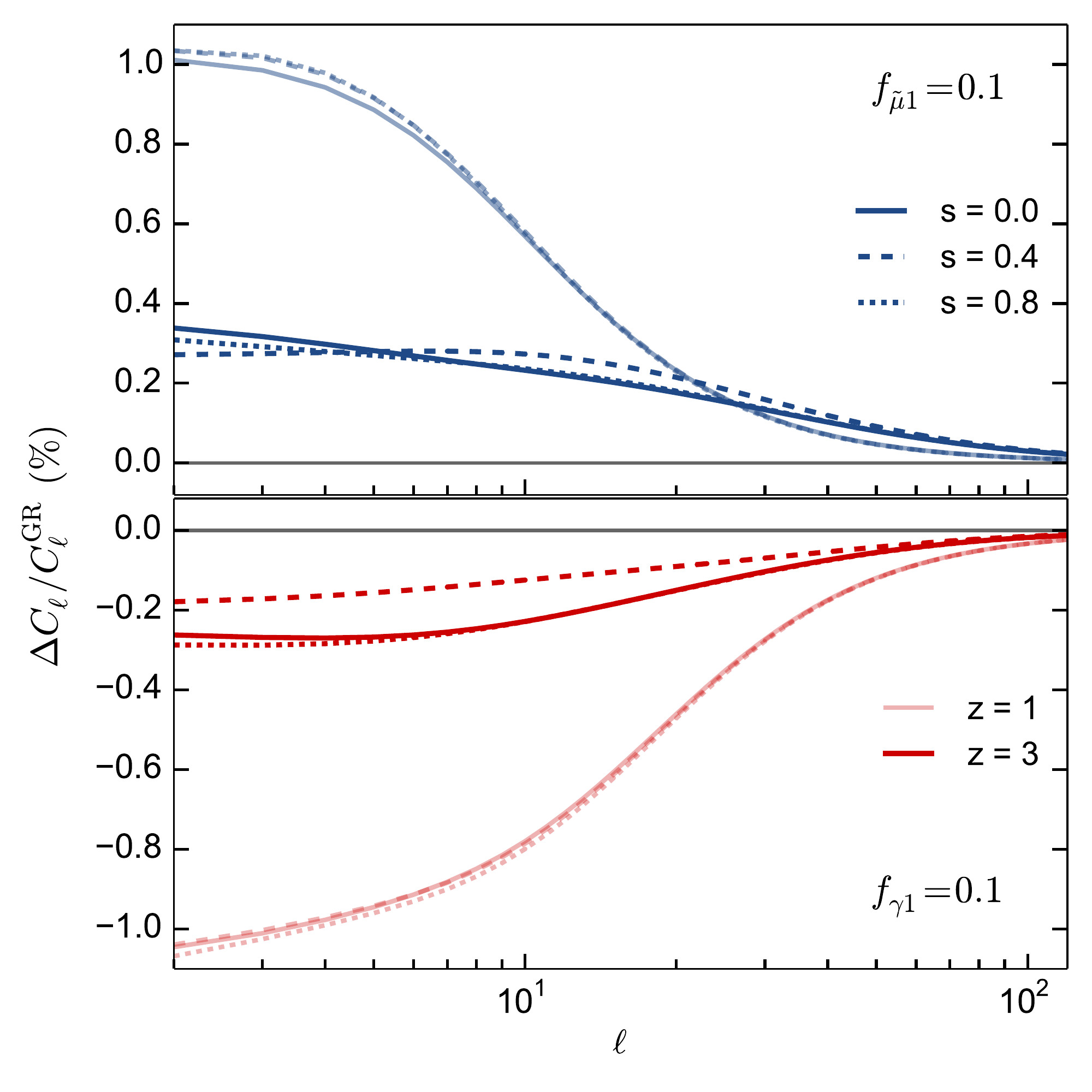}
  \end{minipage}
\caption{Fractional deviation of the number count power spectrum for the $\Lambda$ (left) and CPL-like (right) parameterisations for the $z=1$ and $z=3$ redshift bins (indicated by light/heavy curves), with different values of the magnification bias (solid/dashed/dotted curves). The $s=0.4$ (dashed) curves correspond to the intensity mapping case.}
\label{fig:sbias}
\end{figure*}

\subsection{Calculating relativistic observables in MG}\label{sec:relobservables}

We now wish to obtain the full linear relativistic expression for the number counts, Eq.~(\ref{eq:DeltaN}), in the presence of modifications to GR represented by our Eqs.~(\ref{Poisson_gi}-\ref{jkg}). The derivations in \cite{2011PhRvD..84f3505B} and \cite{2011PhRvD..84d3516C} did not assume the GR field equations, and used only perturbations to purely geometric quantities like the Jacobian and luminosity distance, plus stress-energy conservation. As such, the MG parameterisation described above affects $\Delta_{\rm N}$ only by changing the source terms in Eq.~(\ref{eq:DeltaN}), e.g. by modifying the solutions for $\Phi(k, \eta)$ and $\Psi(k, \eta)$, leaving the form of the equation itself unchanged. This is not completely general -- for example, the assumption of normal stress-energy conservation for matter excludes some gravity theories, like those with non-universal matter couplings \citep{2000PhRvD..62d3511A, 2008PhRvD..77j3003P, 2010JCAP...04..032B}. It is precisely these theories for which the parameterisation of Eqs.~(\ref{Poisson_gi}) and (\ref{slip_gi}) is ill-suited though, and so we will not consider them here. Note that a more general expression, valid for any metric theory of gravity, can be obtained with only relatively minor changes to Eq.~(\ref{eq:DeltaN}) \citep{2014CQGra..31w4002B}.

Up to the assumptions we have made, then, all that is needed to calculate the number count perturbation in MG theories is to modify the GR calculation of Eq.~(\ref{eq:DeltaN}) to use the MG solutions for the source terms $\Phi,\,\Psi,\,V_{\rm M}$, and $D_{\rm M}$. We implement this by modifying the CLASS Boltzmann code\footnote{\url{http://class-code.net}} to calculate the angular power spectrum of $\Delta_{\rm N}$ using the MG perturbation equations, Eqs.~(\ref{Poisson_gi}), (\ref{slip_gi}), and (\ref{eq:newPhidot}). We will consider only late time modifications to GR, which allows us to take GR initial conditions at high redshift, and to leave the CMB calculation unmodified. Details of our implementation are given in Appendix \ref{app:classmod}.

\subsection{MG effects on ultra-large scale observables}

Fig.~\ref{fig:deltacl} shows the fractional difference of the number count angular power spectrum, $C_\ell(z,z)$, from GR, for four different choices of MG ansatz/parameters. The galaxy bias was taken to be constant, $b=2$, and magnification bias and evolution bias were neglected, $s(z) = f^{\rm N}_{\rm evo}(z) = 0$. The onset scale of the modifications in Eqs.~(\ref{mu_ansatz}) and (\ref{gamma_ansatz}) was chosen to be $\lambda = 10$, which ensures that they only become relevant at $k \lesssim 10^{-2}$ Mpc$^{-1}$, i.e. beyond the matter-radiation equality peak in the matter power spectrum. {\corr This approximately marks the maximum scale so far probed by galaxy surveys, as well as the transition away from the sub-horizon, quasi-static regime. One could of course choose a larger value of $\lambda$ to study effects on smaller scales, but we leave this to future work.}

Each of the panels in Fig.~\ref{fig:deltacl} shows the auto-spectra for four redshift bins, $z=\{0.1, 1, 2, 3\}$, all with tophat selection functions of full width $\Delta z = 0.2$. The $z=0.1$ redshift bin consistently shows the smallest deviation from GR, which is straightforward to understand in light of Fig.~\ref{fig:horizon} -- at such small distances from the observer, even the largest angular scales correspond to subhorizon physical scales, where the MG corrections are suppressed in our parameterisation. The picture is more complicated at higher redshift however, as the relative importance of the various contributions to Eq.~(\ref{eq:DeltaN}) varies non-trivially with scale and redshift. For example, the top right panel of Fig.~\ref{fig:OmL_DM} shows that the MG functions ($\tilde{\mu}, \gamma$) grow with redshift for the CPL ansatz, and yet the $z=3$ auto-spectrum deviates from GR {\it less} than the $z=2$ curve. This is not the case for the $f_{\tilde{\mu}\Lambda} = 0.1$ curves, where the $z=3$ curve is larger than the $z=2$ curve for $\ell \le 10$.

{\corr This behaviour can be explained by inspecting the relative size and differing modifications to the various terms in Eq.~(\ref{eq:DeltaN}). The density term grows with time and is generally the dominant contribution to the total $C_\ell$ (see the left panel of Fig.~\ref{fig:deltacl_terms}), but this is not to say that its deviation from GR grows with time, or that it is the dominant contribution to $\Delta C_\ell$. The right-hand panels of Fig.~\ref{fig:deltacl_terms} show the deviation from GR of the $\Lambda$-ansatz power spectra, with the density-only and several non-density terms from Eq.~(\ref{eq:DeltaN}) switched on and off. For $f_{\tilde{\mu}\Lambda} = 0.1$ (left column) and $z = 0.1$ and $1$, the total $\Delta C_\ell$ curve (solid black) follows the behaviour of the density-only curve (solid blue). The non-density terms are generally subdominant to the density term (see the left panel), so while their {\it fractional} deviations can be large (e.g. see the lensing term; short dashed lines, right panels), their absolute contribution to $\Delta C_\ell$ is small at these redshifts. This is not the case at $z=3$, where the total deviation departs from the shape of the density deviation; the non-density terms are more important at this redshift, and the density-only deviation is close to zero, so does not contribute to the total deviation anyway. Note that cross-correlations between the density and non-density terms further complicate this analysis, as do differences between the two ansatzes (e.g. compare the very different behaviours of the density, $D_{\rm M}$, in Fig.~\ref{fig:OmL_DM}). Still, it is clear that the lensing and RSD terms are more important in driving the deviation from GR at high redshifts, making these interesting effects to focus on when searching for possible modifications to GR on ultra-large scales. Other terms also have significant fractional modifications, but are strongly subdominant for all but the lowest multipoles; we study the modifications for the full set of terms in Eq.~(\ref{eq:DeltaN}) in Appendix \ref{app:allterms}.}

Another way of getting a handle on the relative importance of the various terms is to consider different values of the magnification bias, $s(z)$, which multiplies some, but not all, of the terms. Fig.~\ref{fig:sbias} shows the relative deviations for both ansatzes at $z=1$ and $z=3$, for constant $s = \{0.0, 0.4, 0.8\}$. Recall that $s = 0.4$ is the value to use when considering brightness temperature fluctuations (e.g. for intensity mapping), for which lensing, ISW, and some other terms completely cancel. The magnification bias varies significantly between different galaxy surveys (and also with redshift), but is less than $0.4$ for most future surveys \citep{Alonso2015}.

In Fig.~\ref{fig:sbias}, the $z=1$ curves for $f_{\gamma \Lambda} = 0.1$ and $f_{\gamma 1} = 0.1$ (lower panels) barely change with $s$, suggesting that the $s$-dependent terms contribute little to the total deviation. This is not the case at $z=3$, where (e.g.) the integrated lensing contribution is expected to be more important, although the differences are still small. The contrast between the $f_{\tilde{\mu} \Lambda} = 0.1$ and $f_{\tilde{\mu} 1} = 0.1$ curves (upper panels) is more striking, however. For the CPL ansatz, the dependence on $s$ is very weak for both redshift bins, but the $\Lambda$ ansatz is strongly $s$-dependent. {\corr In fact, the $z=3$ curve for $f_{\tilde{\mu} \Lambda} = 0.1$ goes to zero when $s=0.4$, suggesting that the deviation depends almost entirely on the lensing term. This is borne out by Fig.~\ref{fig:deltacl_terms}, which shows that the lensing term has a relatively large magnitude at $z=3$, and sustains the largest fractional modification of any of the other (important) terms.}

\subsection{Observability of deviations from GR}
\label{subsec:observability}

The upshot of the above is that the distinctive dependence of $\Delta C_\ell$ on different MG parameterisations and choices of MG ansatz, the physical parameters of the survey (e.g. bias), and redshift, could in principle be used to differentiate between different MG theories. The deviations from GR are rather small for the MG parameters we have considered here, however, with a typical magnitude of $\sim 0.1\%$ at $\ell \lesssim 10$ for the $\Lambda$ ansatz, and closer to 0.5\% for the CPL ansatz. The deviations increase towards lower $\ell$ for almost all redshift bins, typically becoming important only for $\ell \lesssim 30$, and they all but vanish beyond $\ell = 100$. For reference, the cosmic variance error for a full-sky experiment in a single redshift bin is about 30\% at $\ell=10$, and 10\% at $\ell=100$. The question, then, is whether it is even possible to observe deviations from GR at this level. 

\begin{figure}[t]
\includegraphics[width=\columnwidth]{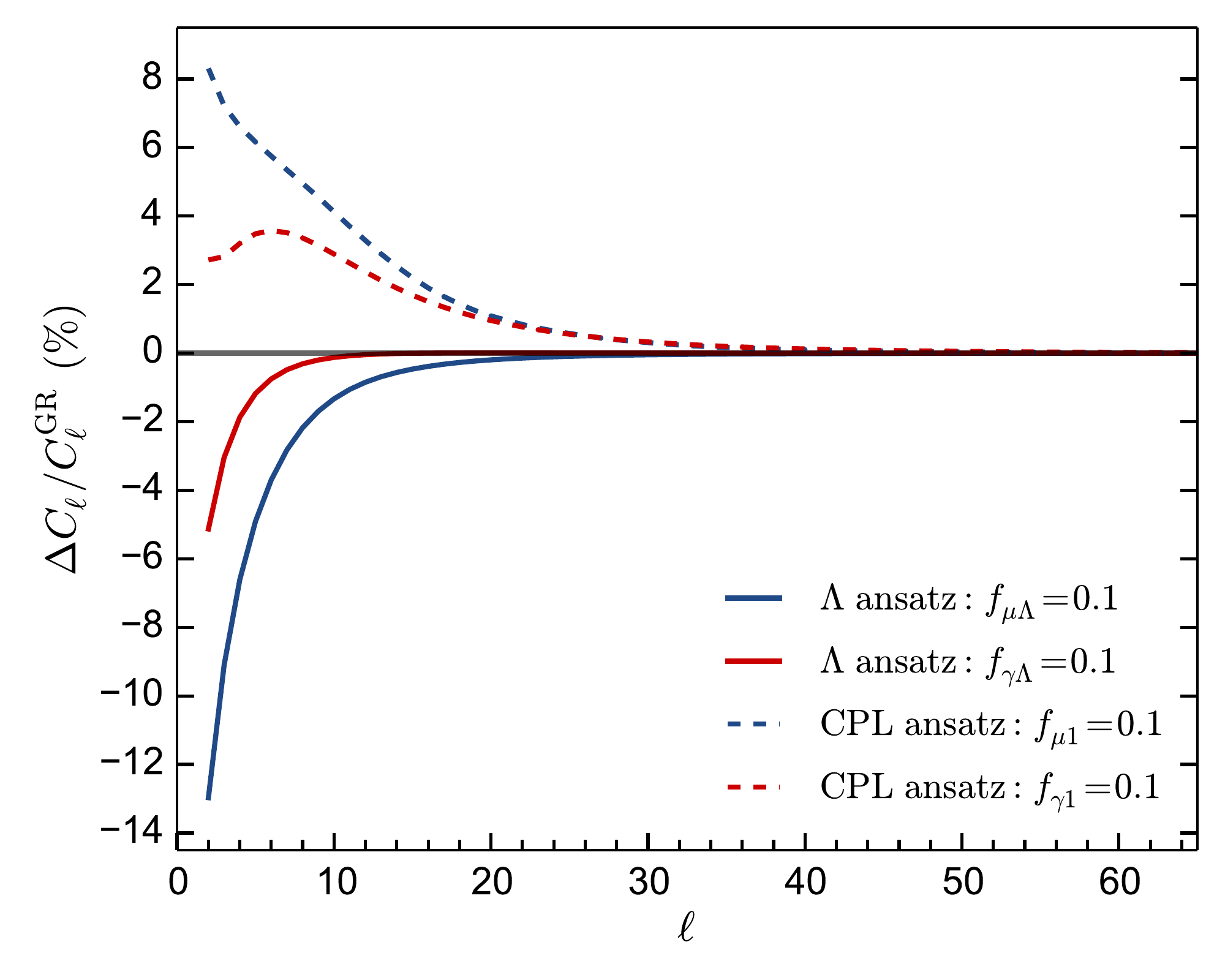}
\caption{Fractional deviation of the (unlensed) CMB temperature power spectrum from its GR behaviour, for both parameterisations. The results are comparable for the lensed temperature spectrum. Broadly speaking, the CPL ansatz shows larger deviations as it affects higher redshifts than the $\Lambda$ ansatz, thereby boosting the integrated Sachs-Wolfe effect.}
\label{fig:cmb}
\end{figure}

It is first prudent to check whether the modifications considered above could have stronger effects on other observables that might be easier to measure. The most obvious place to check is the CMB temperature (TT) power spectrum, which has been measured down to the cosmic variance limit for $\ell \lesssim 1500$ over almost the entire sky \citep{2014A&A...571A..15P}. Our parameterisation leaves early times unmodified by construction, but the observed TT spectrum still depends on late-time physics through secondary anisotropies like lensing and the ISW effect. Fig.~\ref{fig:cmb} shows the deviation of the CMB TT spectrum from GR for the same sets of MG parameters presented in Fig.~\ref{fig:deltacl}. The differences are again negligible for all but the lowest $\ell$ modes, but are substantially larger than for the number count power spectra, weighing in at around 5--10\% at $\ell \lesssim 10$. Nevertheless, they are still well below the cosmic variance error, and so would not be detectable.

While we only observe one CMB, it is possible to observe multiple redshift bins with multiple surveys/tracers, each of which can yield additional information about possible deviations from GR. It is therefore not clear whether the effects shown in Fig.~\ref{fig:cmb} are too small to be observable due to the fundamental limit imposed by cosmic variance. To get a crude estimate of their detectability, we can write the modified number count power spectrum as
\be
C^{\rm MG}_\ell(z_i, z_j) = [1 + A\, g_\ell(z_i)]\, C^{\rm GR}_\ell(z_i, z_j)\, \delta_{i j},
\ee
where we have explicitly neglected cross-correlations, and assume that the deviations can be parameterised by an amplitude parameter, $A$, and a known redshift and angular scaling, $g_\ell(z)$. We can roughly approximate the scaling as being constant with redshift and a step function in $\ell$, such that the deviation is the same in each of $N_{\rm bins}$ redshift bins for all $\ell \le \ell_{\rm max}$. For a noise-free experiment, the single-parameter Fisher matrix for $A$ can be approximated as
\be
F_{AA} \approx \frac{1}{2} f_{\rm sky} N_{\rm bins} \left (\frac{\ell_{\rm max} + 1}{A + 1}\right ) ^2,
\ee
and the relative error on the amplitude is
\be
\frac{\sigma_{A}}{A} \approx \sqrt{\frac{2}{f_{\rm sky} N_{\rm bins}}} \left ( \frac{A^{-1} + 1}{\ell_{\rm max} + 1} \right ).
\ee
For the CPL ansatz, the typical deviation is $A\sim 0.3\%$ out to $\ell_{\rm max} \sim 30$. For a full-sky experiment ($f_{\rm sky} = 1$), one would therefore need $\gtrsim 200$ redshift bins to obtain better than a 100\% error on the amplitude parameter, which is clearly unrealistic (especially when one takes into account the fact that the size of the deviation will change for different redshift bin widths).

\begin{figure}[t]
\hspace{-1em}\includegraphics[width=1.03\columnwidth]{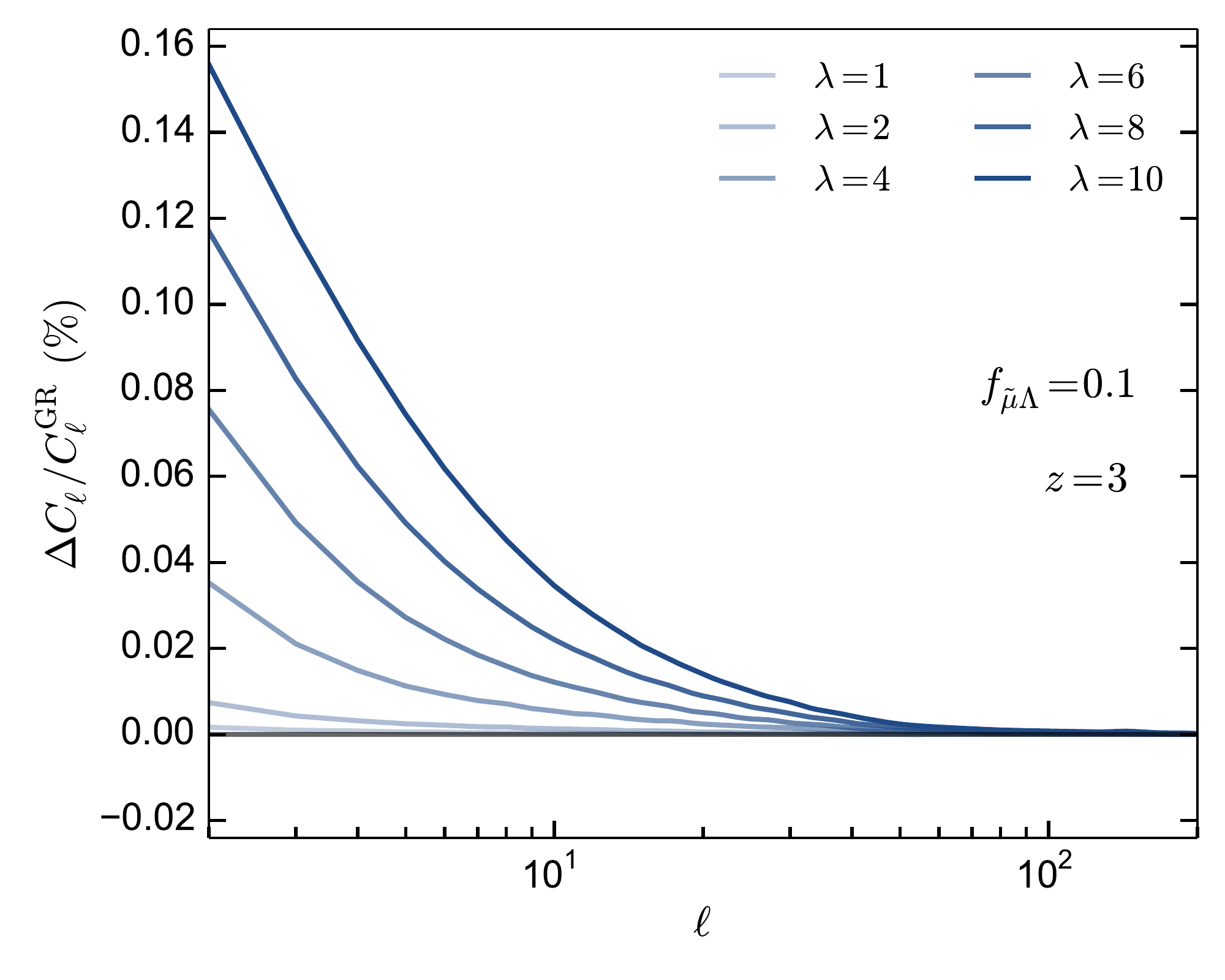}
\caption{Fractional deviation of the number count power spectrum at $z=3$ for the $\Lambda$ parameterisation ($f_{\tilde{\mu}\Lambda} = 0.1$), as a function of the scale of the modifications, $\lambda$.}
\label{fig:omegade_lambda}
\end{figure}

This simple estimate does not give the full picture, as we have concentrated only on one set of survey characteristics, as well as neglecting noise and a number of nuisance parameters. Certain types of survey might exhibit larger deviations, for example due to a different magnification or evolution bias. Since at least some of the terms in Eq.~(\ref{eq:DeltaN}) depend on the particular properties of the tracer population (e.g. $b$, $s$, $f^{\rm N}_{\rm evo}$), it is also possible to use the `multi-tracer' effect \citep{2009PhRvL.102b1302S} to circumvent the cosmic variance limit, as discussed in \cite{2013PhRvD..87j4019L}. Multi-tracer analyses also appear to be necessary to detect other ultra-large scale effects, like the scale-dependent bias due to primordial non-Gaussianity \citep{2009JCAP...10..007M, 2008PhRvD..77l3514D, Ferramacho:2014pua}. Indeed, most of the relativistic terms in the number count spectra are undetectable in single-tracer analyses \citep{2012PhRvD..86f3514Y, Alonso2015}.

We have also considered only a single set of MG parameters, and certain viable theories might predict significantly larger (or smaller) effects. We have argued that modifications around the Hubble scale ($\lambda \sim 1$) with a similar redshift scaling to dark energy are in some sense `natural' if the MG theory has been invoked to explain cosmic acceleration, so the choices we made in this regard should be somewhat representative. We did make relatively arbitrary choices of the amplitude parameters, $f_{\tilde{\mu},\gamma} = 0.1$, and fixed $\lambda$ to a slightly higher value of 10, however, which may not be so typical. Fig.~\ref{fig:omegade_lambda} shows the effect of changing $\lambda$, but a more detailed investigation of the parameter space is necessary to draw more solid conclusions (e.g. see \cite{2013PhRvD..87j4019L} for the effects in some example theories). Nevertheless, it seems likely that the deviations of the observables will be small, and thus difficult to detect, in most cases.

\section{Conclusions}
\label{sec:conclusions}
An impasse is emerging in the use of conventional galaxy surveys to test ideas about cosmic acceleration. As constraints from the linear, subhorizon regime are cycled back into MG models, theorists will likely begin to focus on theories whose phenomenology becomes manifest only outside the linear regime. Experiments are pushing towards high-accuracy measurements of smaller scales, but interpretation of the data there remains challenging due to nonlinearity and poorly-understood baryonic physics, even for $\Lambda$CDM. Tests of gravity on these scales rely heavily on computer simulations, and while rapid progress is being made \citep{Winther2014, Barreira2014, Falck2015}, these remain resource-intensive and difficult to validate. 
 
Given this situation, it is logical to consider exploring the \textit{other} end of the distance ladder. Although accurate measurements of horizon-scale modes are, experimentally, slightly further off than small-scale advances, they could ultimately prove more useful for shaping ideas about alternative gravity theories. If a theory's effects are ruled out on horizon scales, then it becomes questionable whether it can be of any use in driving cosmic acceleration, irrespective of small-scale phenomenology. An additional advantage is that, from the perspective of using effective field theory (EFT) to explore MG models, large cosmological scales are safely in the low-energy, linear regime. This guarantees that one has a reliable theoretical understanding of the EFT, and concrete predictions can be made.
  
 This is not to say that ultra large-scale cosmology will not face its own challenges. A particular obstacle is that any new horizon-scale physics emerging at late cosmological times only imprints its effects on a small number of angular multipoles. As we have shown in this paper, deviations from the GR number count power spectra are $\lesssim 1\%$ for $\sim 10\%$ corrections to the Poisson equation and slip relation, assuming that the MG effects occur on scales just above current observational reach. Deviations of this magnitude will be dominated by cosmic variance errors, although there is some hope that this could be ameliorated through the use of multi-tracer techniques. It is also possible that specific gravity theories may be found that behave very differently from the generalised parameterisation used here, resulting in more readily detectable deviations.
 
 The contribution of relativistic corrections to the angular spectrum of number counts leads to interesting scale- and redshift-dependence of the MG deviations. {\corr Some of the terms (e.g. lensing) are also modified more strongly than others, which is potentially very useful if they can be separated out. It will be a struggle to isolate most of the relativistic corrections from the contribution of the matter density perturbation, however, even with the SKA \citep{2013JCAP...02..044M, Alonso2015}. The best prospect is likely to be a multi-tracer analysis; the cross-correlation of an SKA 21cm intensity mapping survey with photometric galaxy redshifts from LSST is sufficient for $\sim\!\!5-20\sigma$ detections of the amplitude of the lensing term and the combination of the other relativistic terms, for example \citep{Alonso:2015sfa}. If we find in future that the relativistic effects are more clearly discernible than expected, then this could also be strong evidence that there is some large-scale modification to gravity boosting these terms.}
 
One could draw a parallel between our investigation of large-scale MG phenomenology and that of direct detection of (non-primordial) gravitational waves (GWs). Though an initial detection of GWs has yet to be made, the effects of modified gravity on compact object inspirals are investigated (e.g. \citealt{Berti2013}), and model-independent tools for GW analysis are constructed \citep{Cornish2011} in preparation for a future era where GW detection is well-established. In a similar manner, this paper has presented an exploration of how simple changes to the linearised gravitational field equations propagate through to shifts in large-scale matter density perturbations and related observables. As observational constraints on scale-dependent growth continue to tighten,  we are forced to `scan the horizon' in this way if modified gravity is to remain a viable (but ultimately detectable) contender for explaining cosmic acceleration.


\acknowledgments{\it Acknowledgements ---} We are grateful to  D. Alonso, C.~Bonvin, P.~G. Ferreira, R. Maartens, and J. Noller for a number of useful discussions. TB is supported by All Souls College, Oxford. PB is supported by European Research Council grant StG2010-257080.

\begin{appendix}

\section{Large-scale Deviation of $\delta D_{\rm M}$}
\label{app:DM}

 Here we provide a few more details on the linearised deviations from $\Lambda$CDM+GR on superhorizon scales that were discussed in \S\ref{subsec:limits}. First we convert the independent variable to $x=\ln a$, denoting derivatives with respect to $x$ by a prime. Eq.~(\ref{deltaY}) becomes
\begin{align}
\delta {Y\pp}+\delta {Y\p}\left(2+\frac{\Hu\p}{\Hu}\right) -{Y\p}_{\rm GR}\delta\gamma\simeq 0 \label{app_ddY}.
\end{align}
It is helpful to introduce the variable $\delta X = {\delta{Y\p}_{\rm M}}/{{Y\p}_{\rm M}^{\rm GR}}$. Making use of Eq.~(\ref{YULSGR}), we have
\begin{align}
\delta  X\p &= \frac{\delta{Y\pp}_{\rm M}}{{Y\p}_{\rm M}^{\rm GR}}-\frac{{Y\pp}_{\rm M}^{\rm GR}}{{Y\p}_{\rm M}^{\rm GR}}\,\delta X\\
&= \frac{\delta{Y\pp}_{\rm M}}{{Y\p}_{\rm M}^{\rm GR}}+\left(2+\frac{\Hu\p}{\Hu}\right)\,\delta X,
\end{align}
so that Eq.~(\ref{app_ddY}) becomes simply
\begin{align}
\delta X\p - \delta\gamma \simeq 0.
\end{align}
Integrating this in conformal time and substituting into Eq.~(\ref{deltadeltaM}), we then obtain
\begin{align}
\frac{\delta D_{\rm M}}{D_{\rm M}^{\rm GR}}&\simeq \int\left(\Hu \,\delta\gamma\right)d\eta-\left(\delta\tilde\mu+\delta\gamma\right)\label{app_ddM}.
\end{align}

\section{Conservation of $\cR_{\mathrm{co}}$}
\label{app:Rconsv}

In this appendix, we show that the comoving curvature perturbation is conserved on superhorizon scales in the presence of our modifications to the linearised field equations.  
The comoving curvature perturbation at late times, when pressureless matter is dominant, is
\begin{align}
\label{CNR}
\coR=-\hat\Phi-\frac{\Hu}{k}V_{\rm M}.
\end{align}
Differentiating this and using the second of Eqs.~(\ref{consv}) gives
\begin{align}
\dot{\cR}_{\mathrm{co}}&=-(\dot{\hat\Phi}+\Hu\hat\Psi)+\frac{V_{\rm M}}{k}(\Hu^2-\dot\Hu).
\label{Rprime1}
\end{align}
After some straightforward manipulation, one can eliminate $V_{\rm M}$ in favour of metric potentials. We differentiate our parameterised Poisson equation (\ref{Poisson_gi}) and use the definition of $Y_{\rm M}$ and the fluid conservation equations to obtain:
\begin{align}
-2k^2\left(\dot{\hat\Psi}-C\hat\Psi\right)&=8\pi G a^2 \rho_{\rm M}\mt\,\dot{D}_{\rm M}\\
&=8\pi G a^2 \rho_{\rm M}\mt\Bigg[-k V_{\rm M}+3\left(\dot{\hat\Phi}+\Hu\Psi\right) +3\frac{V_{\rm M}}{k}\left(\dot\Hu-\Hu^2\right)\Bigg],
\end{align}
where $C ={d/d\eta}\left[\ln(a^2\rho_{\rm M} \mt\right]$. Rearranging, we obtain
\begin{align}
\frac{V_{\rm M}}{k}&=\frac{1}{\left[3(\dot\Hu-\Hu^2)-k^2\right]}\Bigg[\frac{-2k^2(\dot{\hat\Psi}-C\hat\Psi)}{8\pi Ga^2\rho_{\rm M}\mt}- 3(\dot{\hat\Phi}+\Hu\hat\Psi)\Bigg].
\label{vsub}
\end{align}
Substituting this into Eq.~(\ref{Rprime1}), we finally get
\begin{align}
\dot{\cR}_{\mathrm{co}} =-(\dot{\hat\Phi}+\Hu\hat\Psi) + \frac{(\dot\Hu-\Hu^2)}{\left[3(\dot\Hu-\Hu^2)-k^2\right]}\Bigg[\frac{2k^2\left(\dot{\hat\Psi}-C\hat\Psi\right)}{8\pi Ga^2\rho_{\rm M}\mt}+3\left(\dot{\hat\Phi}+\Hu\hat\Psi\right)\Bigg].
\label{Rprime2}
\end{align}
In the limit $k^2\rightarrow 0$, the RHS of the above expression vanishes identically if we impose the reasonable restriction that $\mt$ does not tend to zero on ultra-large scales. The variable $\coR$ is therefore conserved on superhorizon scales in our generalised scenario, in agreement with the findings of \cite{Wands:2000jd} and \cite{Bertschinger:2006ht}.

Now that we know $\dot{\cR}_{\mathrm{co}}\rightarrow 0$ on large scales, Eq.~(\ref{ddPhiULS}) is straightforwardly obtained by substituting the $k^2\rightarrow 0$ limit of Eq.~(\ref{vsub}) and its time derivative into (\ref{Rprime1}).

{\corr 
\section{Term-by-term deviations from GR}
\label{app:allterms}

\begin{figure}[t]
\centering{
\includegraphics[width=0.86\columnwidth]{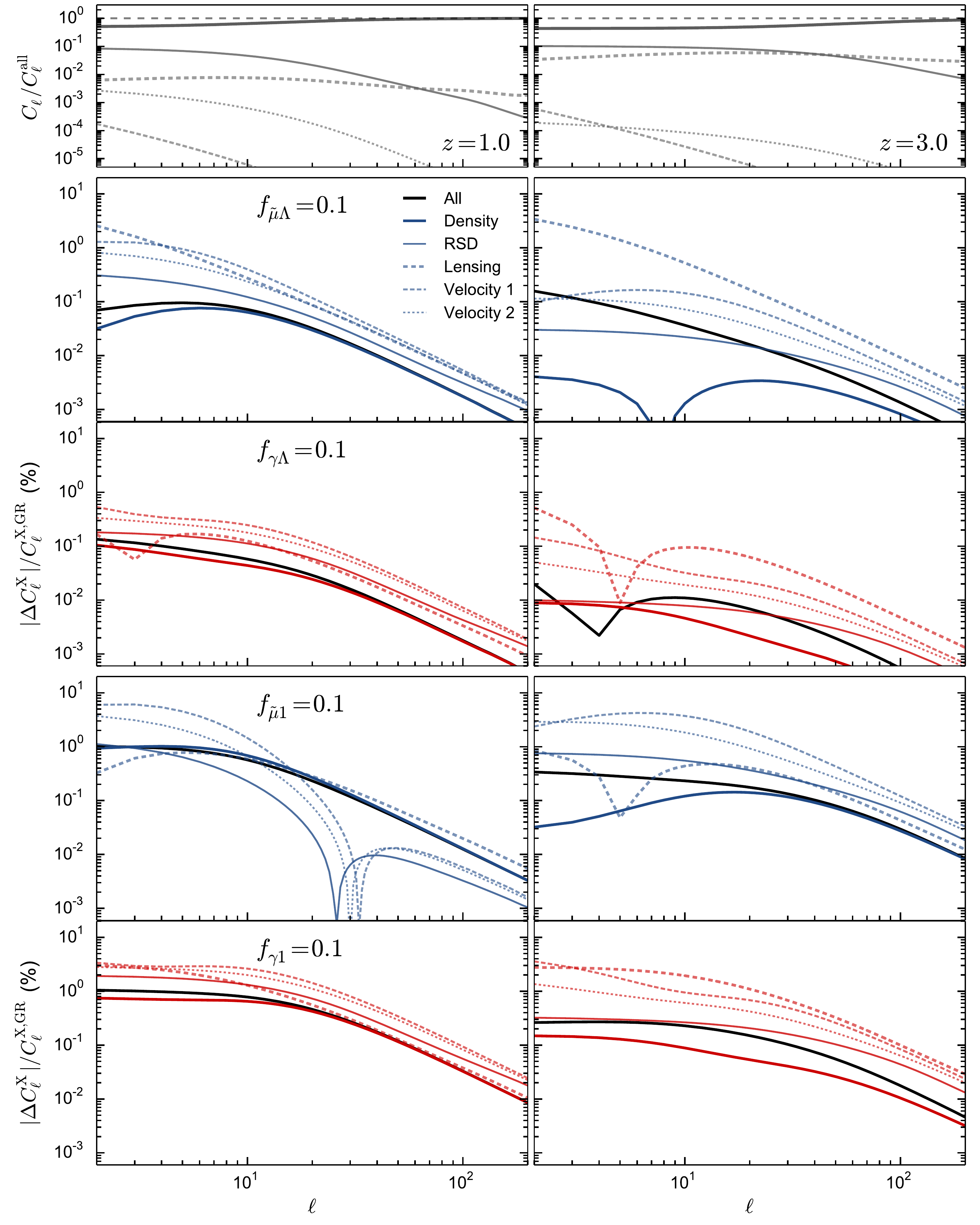}
}
\caption{Top panels: Fractional contribution of various terms from Eq.~(\ref{eq:DeltaN}) to the total angular power spectrum, $C^{\rm all}_\ell$, for the $z=1$ (left column) and $z=3$ (right column) redshift bins. Middle and lower panels: Relative deviation of each of the terms from its GR value, for 
the $\Lambda$-ansatz (middle) and CPL-like ansatz (lower), again for the $z=1$ and $z=3$ redshift bins. The solid black line shows the total deviation from GR, $\Delta C_\ell^{\rm all} / C^{\rm all, GR}_\ell$. Note that we have taken the absolute value of the relative deviation (some of the deviations are negative).}
\label{fig:allterms_leading}
\end{figure}

Figs.~\ref{fig:allterms_leading} and \ref{fig:allterms_potentials} show the fractional deviation of the number count power spectrum for individual terms in Eq.~(\ref{eq:DeltaN}). Results are shown for two redshift bins, for each of the MG ansatzes considered above. The top panels show the absolute size of each of the terms in the GR case. We follow the naming convention for those terms given by \cite{Alonso2015}; see Section 2.1.5. of that paper for the explicit expressions.

\begin{figure}[t]
\centering{
\includegraphics[width=0.86\columnwidth]{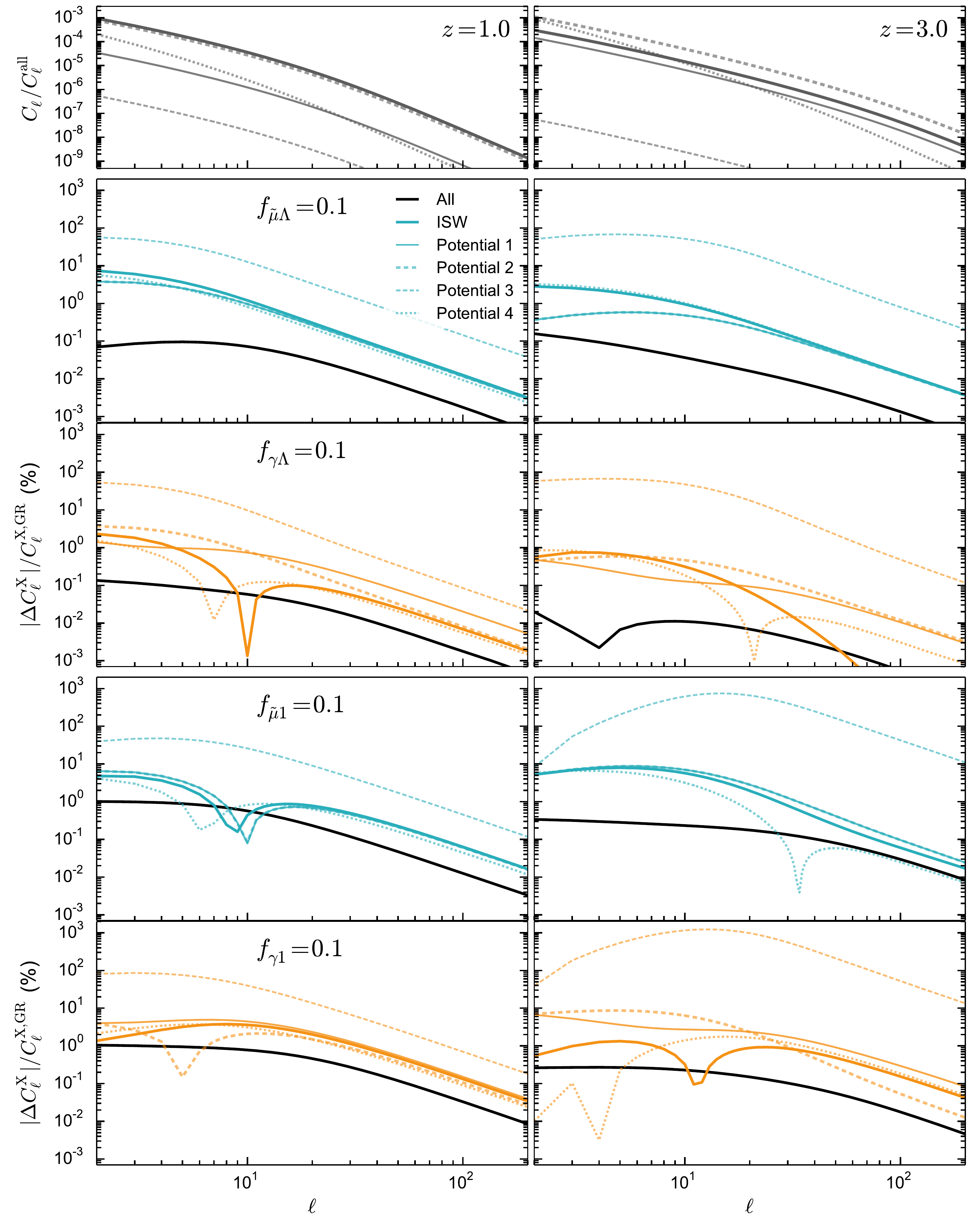}
}
\caption{The same as Fig.~\ref{fig:allterms_leading}, but for the remaining set of terms from Eq.~(\ref{eq:DeltaN}). The various spikes are due to $\Delta C_\ell$ crossing zero.}
\label{fig:allterms_potentials}
\end{figure}

Note that we only plot the autocorrelation for each term, e.g. the `Density' curve shows the density-density part of the number count power spectrum only. We have neglected to show cross-terms here, even though (e.g.) the density-RSD term is known to be significant.

We briefly note three features of these figures. First, it is clear that the density term always dominates the total power spectrum at these redshifts, with the lensing and RSD terms being the only other significant contributions. All other terms are significantly smaller (except for at the very lowest $\ell \lesssim 5$), and decay rapidly with $\ell$.

The deviation of the density term from GR does not necessarily dominate the total deviation, however, as in some cases $\Delta C_\ell^{\rm Dens}$ is very small. For example, for $f_{\mt \Lambda} = 0.1$ at $z=3$ (Fig.~\ref{fig:allterms_leading}, top right deviation panel), the density term never deviates from GR by more than $5 \times 10^{-3}\%$. While the absolute magnitude of the lensing term is around an order of magnitude smaller than the density term, its fractional deviation is 3 orders of magnitude larger. It therefore dominates the total fractional deviation, as can be seen from the similar shapes of the `Lensing' and `All' curves.

Finally, figure Fig.~\ref{fig:allterms_potentials} shows that many of the subdominant `potential' terms, while very small in absolute terms, are modified by the largest fractional amount. For $f_{\gamma\Lambda}=0.1$ at $z=3$, for example, the `Potential 3' term (the smallest of all the terms, $\propto \aH^{-1} \dot\Phi$) is modified by almost 100\% at $\ell \approx 10$. While this dramatic {\it fractional} modification is unobservable in this context, it may be worth looking for other observables that are more sensitive to the absolute value of this term.
} 

\vfill

\section{Modifications to CLASS}
\label{app:classmod}

The angular power spectrum calculations in this paper were performed using a modified version of CLASS 2.4.2, a publicly available Boltzmann code \citep{2011JCAP...07..034B}. The code is written in such a way that only a few modifications to the source terms of the perturbation evolution equations are needed -- everything else is sufficiently general to accommodate our MG corrections at late time. We introduced our changes into the Newtonian gauge part of the code only (CLASS can use both Newtonian and Synchronous gauge), and allowed our modifications to switch on only at late times, $z \le 6$, falling back to the default CLASS code at higher redshifts.

The metric potential $\Phi$ is evolved using an equation that depends on the GR field equations by default. We replaced this with an expression derived by differentiating the modified Poisson equation with respect to conformal time, and then substituting Eq.~(\ref{consv}) to get
\be \label{eq:newPhidot}
\dot{\Phi} = \left ( 1 + \frac{9}{2}\frac{\aH^2}{k^2} \Omega_{\rm M} \mt \gamma \right )^{-1} \left [ \Phi \left ( \frac{\dot{\gamma}}{\gamma} + \frac{\dot{\tilde{\mu}}}{\tilde{\mu}} - \aH \right )  + \frac{9}{2}\frac{\aH^2}{k^2} \Omega_{\rm M} \mt \gamma V \left ( \frac{k}{3} + (\aH^2 - \dot{\aH})/k\right ) - \frac{9}{2}\frac{\aH^2}{k^2} \Omega_{\rm M} \mt \gamma \Psi \,\aH \right ],
\ee
which is valid at late times for the standard fluid stress-energy sources. The other metric potential, $\Psi$, is derived from the solution for $\Phi$ using the slip relation, which we also replaced with the modified version. We validated our modifications by confirming that the code gives the same results as the unmodified CLASS code in the GR limit, $\tilde{\mu}=\gamma=1$. We also wrote an independent Python implementation of the MG evolution equations, and confirmed that the results agree with the evolution calculated by the modified CLASS code.

The precision settings of the CLASS integrator must be increased considerably beyond their defaults to obtain sufficient accuracy for the power spectrum calculations (i.e. to obtain an accuracy much better that the typical $\sim\! 10^{-3}$ differences between GR and MG spectra found above). This can increase running time considerably, but is necessary to avoid the results being dominated by numerical noise. {\corr We increased the Bessel function sampling to \texttt{selection\_sampling\_bessel=3.0}, reduced the Fourier-space step size in the transfer function calculation to \texttt{k\_step\_trans\_scalars=0.3} to improve the resolution of the integrator, and set the tolerance of the perturbation solver to \texttt{tol\_perturb\_integration=1e-7}. We found that these choices give sufficiently stable results that do not change appreciably with higher precision settings. We also set \texttt{k\_scalar\_max\_tau0\_over\_l\_max = 4} and switched off the Limber approximation at all $\ell$ values, which improves the precision at higher $\ell$.}

Our modifications to CLASS are publicly available at \url{https://gitlab.com/philbull/mgclass}.

\end{appendix}

\bibliography{mg_uls}
\bibliographystyle{hapj}

\end{document}